\documentclass[longauth]{aa}  
\usepackage[colorlinks=true,linkcolor=blue,citecolor=blue]{hyperref}
\usepackage{graphicx}
\usepackage{float}
\usepackage{natbib}
\usepackage{xcolor}
\bibpunct{(}{)}{;}{a}{}{,} 
\usepackage{txfonts}
\DeclareMathAlphabet{\mathpzc}{OT1}{pzc}{m}{it}
\begin{document}

   \title{The GAPS programme at TNG \thanks{Based on observations made with the Italian Telescopio Nazionale
{\it Galileo} (TNG) operated by the Fundaci\'on Galileo Galilei (FGG) of the
Istituto Nazionale di Astrofisica (INAF) at the
Observatorio del Roque de los Muchachos (La Palma, Canary Islands, Spain).}}

\subtitle{LX. Atmospheric characterisation of KELT-9\,b via single-line analysis: Detection of six H {\sc i} Balmer lines, Na {\sc i},  Ca {\sc i}, Ca {\sc ii}, Fe {\sc i}, Fe {\sc ii}, Mg {\sc i}, Ti {\sc ii}, Sc {\sc ii}, and Cr {\sc ii}.}

   \author{M. C. D'Arpa
          \inst{1,}\inst{2}, 
          A. Saba\inst{3}, 
          F. Borsa\inst{4}, 
          L. Fossati\inst{5}, 
          G. Micela\inst{1}, 
          C. Di Maio\inst{1}, 
          M.~Stangret\inst{6}, 
          G. Tripodo\inst{2},         
          L.~Affer \inst{1},  
          A.~S. Bonomo \inst{7},
          S.~Benatti \inst{1}, 
          M.~Brogi \inst{8,7}, 
          V.~Fardella \inst{1, 2}, 
          A.~F.~Lanza\inst{9}, 
          G.~Guilluy \inst{7}, 
          J.~Maldonado \inst{1}, 
          G.~Mantovan \inst{6,10}, 
          V.~Nascimbeni \inst{6}, 
          L.~Pino \inst{11}, 
          G.~Scandariato \inst{9}, 
          D.~Sicilia \inst{9}, 
          A.~Sozzetti\inst{7}, 
          R.~Spinelli \inst{1}, 
          G.~Andreuzzi \inst{12,13}, 
          A.~Bignamini \inst{14}, 
          R.~Claudi\inst{6}, 
          S.~Desidera \inst{6}, 
          A.~Ghedina \inst{13}, 
          C.~Knapic \inst{14},     
          V.~Lorenzi \inst{12,13}
          }
          
   \institute{INAF -- Palermo Astronomical Observatory, Piazza del Parlamento, 1, 90134 Palermo, Italy\\  
              \email{mattia.darpa@inaf.it} 
         \and
             University of Palermo, Department of Physics and Chemistry “Emilio Segrè", Via Archirafi 36, Palermo, Italy 
        \and
             Department of Physics and Astronomy, University College London, Gower Street, WC1E 6BT London, United Kingdom 
        \and 
        INAF -- Osservatorio Astronomico di Brera, Via E. Bianchi 46, 23807 Merate (LC), Italy 
        \and 
        Space Research Institute, Austrian Academy of Sciences, Schmiedlstrasse 6, 8042 Graz, Austria 
        \and
        INAF -- Osservatorio Astronomico di Padova, Vicolo dell'Osservatorio 5, 35122, Padova, Italy 
        \and
        INAF -- Osservatorio Astrofisico di Torino, Via Osservatorio 20, 10025, Pino Torinese, Italy 
        \and
        Dipartimento di Fisica, Universitá degli Studi di Torino, Via Pietro Giuria 1, I
        10125 Torino, Italy 
        \and
        INAF -- Osservatorio Astrofisico di Catania, Via S. Sofia 78, I-95123, Catania, Italy 
        \and
        Dipartimento di Fisica e Astronomia ``Galileo Galilei'', Università di Padova, Vicolo dell'Osservatorio 3, IT-35122, Padova, Italy 
        \and
        INAF -- Osservatorio Astrofisico di Arcetri, Largo E. Fermi 5, 50125, Firenze, Italy 
        \and
        Fundación Galileo Galilei-INAF, Rambla José Ana Fernandez Pérez 7, 38712 Breña Baja, TF, Spain 
        \and
        INAF - Osservatorio Astronomico di Roma, Via Frascati 33, 00078 Monte Porzio Catone, Italy 
        \and
        INAF – Osservatorio Astronomico di Trieste, via Tiepolo 11, 34143 Trieste 
}

   \date{-}

 
  \abstract
  {}
   {We analysed six primary transits of the ultra-hot Jupiter KELT-9\,b obtained with the HARPS-N high-resolution spectrograph in the context of the Global Architecture of Planetary Systems (GAPS2) project, to characterise the atmosphere via single-line analysis.}
   {We extracted the transmission spectrum of each individual line by comparing the master out-of-transit spectrum with the in-transit spectra and computing the weighted average of the tomography in the planet reference frame. We corrected for the centre-to-limb variation and the Rossiter-McLaughlin effect by modelling the region of the star disc obscured by the planet during the transit and subtracting it from the master-out spectrum.}
   {We detected all six observable lines of the Balmer series within the HARPS-N wavelength range, from H$\alpha$ to H$\zeta$, with a significance exceeding 5$\sigma$. We also focussed on metal species, detecting Na {\sc i}, Ca {\sc i}, Ca {\sc ii}, Fe {\sc i}, Fe {\sc ii}, Mg {\sc i}, Ti {\sc ii}, Sc {\sc ii}, and Cr {\sc ii} lines. This is the first detection in the atmosphere of an exoplanet of H$\epsilon$ and H$\zeta$ lines, as well as of individual lines of Sc {\sc ii} and Cr {\sc ii}. Our detections are supported by a comparison with published synthetic transmission spectra of KELT-9b obtained accounting for non-local thermodynamic equilibrium effects. The results underline the presence of a systematic blueshift due to night-side to day-side winds.}
   {The single-line analysis allowed us not only to assess the presence of atomic species in the atmosphere of KELT-9 b, but also to further characterise the local stratification of the atmosphere. Coupling the height distribution of the detected species with the velocity shift retrieved, we acknowledged the height distribution of night-side to day-side winds. Moreover, the study of the rotational broadening of the different species supports the prediction of a tidally locked planet rotating as a rigid body.}


    \titlerunning{Atmospheric characterisation of KELT-9\,b via single-line analysis}
    \authorrunning{M.C.D'Arpa et al.}
   \maketitle
%
\begin{table*}[ht]
\caption{HARPS-N observing log.}
\centering
\begin{tabular}{cccccccc}
\hline
Night                & Date                 & Programme              &  P.I.   & Spectra (In/Out of transit)         & Mean S/N \tablefootmark{a} & 
       Exposure time [s] & Airmass \\ \hline
1                    & 2017-07-31           & A35DDT4              & Ehrenreich & 49 (24/25)            & 137                  & 600   & 1.019 - 1.730  \\
2                    & 2018-06-10           & GAPS                 &  Micela  & 36 (23/13)            & 123                  & 600   & 1.019 - 2.164   \\
3                    & 2018-07-20           & OPT18A\_38            &  Ehrenreich   & 46 (23/23)            & 120                  & 600  & 1.019 - 1.675  \\
4                    & 2018-07-23           & GAPS                 &  Micela  & 68 (44/24)            & 79                   & 300   & 1.019 - 1.535  \\
5                    & 2018-09-01           & GAPS                 &  Micela  & 57 (44/13)            & 99                   & 300   & 1.019 -  1.800  \\
6                    & 2018-09-04           & GAPS                 &  Micela  & 64 (44/20)            & 87                   & 300   & 1.019 - 1.426   \\ \hline\\
\label{tab:observations}
\end{tabular}
\tablefoot{
\tablefoottext{a}{The mean S/N has been computed as the average of all the orders of all the spectra for each night. We show the S/N as a function of wavelength in Fig~A.1.}}
\end{table*}

\section{Introduction}
The wide sample of more than 5000 detected exoplanets includes a broad selection of different alien worlds spanning from cold super-Earths to extremely hot Jupiter-sized planets. This latter category hosts planets with equilibrium temperatures above 2000 K, which thanks to their extended atmospheres are optimal targets for chemical characterisation.
Among the ultra-hot Jupiters (UHJs), we locate KELT-9\,b, the hottest planet discovered to date \citep{2017Natur.546..514G}. The exoplanet is in a nearly polar ($\lambda$ = 85.78 deg), short-period (1.48 days) orbit at a separation of about 0.03 AU from its host star. Also known as HD195689, the B9.5–A0 host star features an effective temperature in the range of 10,000 K, a radius of $R_{*}$=2.36 R$_\odot$, and a mass ($M_{*}$=2.52 M$_\odot$) more than twice that of the Sun \citep{2017Natur.546..514G}. The tremendous stellar irradiation experienced by KELT-9\,b leads its day-side equilibrium temperature to reach about 4600 K \citep{2017Natur.546..514G}, making this planet hotter than most stars. Hence, its scorching temperature induces the atmosphere to inflate substantially, making this planet an ideal target for transmission spectroscopy.

 In the context of exoplanet characterisation, high-resolution spectroscopy is one of the most powerful methods of investigating the atomic and molecular atmospheric make-up.  Compared to their space-based counterparts, ground-based spectrographs are able to achieve a higher resolving power, such as the $\mathpzc{R}$$\sim$115,000 of HARPS-N \citep{2012SPIE.8446E..1VC}. At such resolutions, individual lines can be resolved, breaking the degeneracy induced by broad spectral features in low-resolution spectra \citep{2018arXiv180604617B}.

A wide variety of metallic species, both ionised and neutral, have been identified in the atmosphere of KELT-9\,b. The cross-correlation technique led to the detection of Fe {\sc i}, Fe {\sc ii}, Ti {\sc ii} \citep{hoeijmakers2018atomic}, Na {\sc i}, Cr {\sc ii}, Sc {\sc ii} and Y {\sc ii} signatures \citep{hoeijmakers2019spectral}, Ca {\sc i}, Cr {\sc i}, Ni {\sc i}, Sr {\sc ii}, and Tb {\sc ii} at the 5$\sigma$ level, and Ti {\sc i}, V {\sc i}, and Ba {\sc ii} above the 3$\sigma$ level \citep{borsato2023mantis}. Furthermore, line studies of its primary transit unveiled the presence of O {\sc i} \citep{borsa_ox_nature}, Mg {\sc i} and Fe {\sc ii} \citep{cauley2019atmospheric}, H$\alpha$ \citep{yan2018extended}, ionised calcium (i.e. H\&K doublet and near-infrared triplet) \citep{yan2019ionized, turner2020detection},
the hydrogen Balmer series up to H$\delta$ \citep{yan2018extended, cauley2019atmospheric, turner2020detection, wyttenbach2020mass}, and the Paschen $\beta$ line \citep{pas_beta}.

In this study, we combine six primary transits of KELT-9\,b observed with the HARPS-N instrument mounted on the \textit{Telescopio Nazionale Galileo} (TNG). By targeting the hydrogen Balmer line series, we aim to identify and characterise all the single H {\sc i} lines up to H$\zeta$, after a careful correction of the strong Doppler shadow effect generated by the geometry of the transit. The same framework adopted for the Balmer series has been used to analyse several metal lines looking for both neutral and ionised species. Finally, we aim to compare the observed features with published theoretical transmission spectra of KELT-9b computed accounting for non-local thermodynamic equilibrium (NLTE) effects that have been shown to reproduce well past observations of the H$\alpha$ and H$\beta$ lines \citep{fossati_models}.

The paper is organised as follows. In Section \ref{sec:observations}, we describe the observational data. Section \ref{sec:methods} illustrates the methodology employed to analyse the data, including the telluric correction, the Doppler shadow removal, and the transit spectrum construction. In Section \ref{sec:results}, we report the results we obtained via the single-line analysis technique, while we compare our transmission spectra with models computed accounting for NLTE effects in Section \ref{sec:comp_models}. We discuss the implications of our study in Section \ref{sec:discussion}, before giving a short summary highlighting the main conclusions in Section \ref{sec:conclusions}.

\section{Observations} \label{sec:observations}
We analysed a total of six KELT-9\,b primary transits, observed from July 2017 to September 2018 with HARPS-N at the TNG. Four of these transits were acquired in the context of the long-term observing programme
at the TNG telescope ‘GAPS2: the origin of planetary systems’ – awarded to the Italian Global Architecture of Planetary System (GAPS) Collaboration \citep[P.I Micela; see][]{2022A&A...665A.104G}, while the two remaining ones were collected by other programmes (P.I. Ehrenreich). HARPS-N, being the northern-hemisphere's twin of ESO's HARPS/3.6 m telescope, is a fibre-fed cross-dispersed echelle spectrograph that covers the 3830--6900 $\AA$ spectral range at an average resolution of $\mathpzc{R}$ $\sim$ 115,000. For each night of observation, the exposure time per spectrum was set to either 600 s or 300 s, leading to an average signal-to-noise ratio (S/N), averaged over all the spectral orders, of 126 and 88, respectively. In addition to the spectra taken in transit, a number of out-of-transit observations were also acquired. These act as a baseline for the stellar flux and allow one to derive the master-out spectrum, which includes only the stellar contribution (see Section \ref{sec:methods}). A complete overview of the observations, including the date, programme number, PI, number of in- and out-of-transit spectra, average S/N, and airmass range, is given in Table \ref{tab:observations}. We also show the S/N as a function of wavelength in Fig. A.1\footnote{See Data availability section after the acknowledgements}. The night reports for the different nights do not give any important observing conditions to remark with a seeing below 1" expect for Night 4. For the analysis of H$\zeta$, we had to discard eight spectra (out of 202 in-transit spectra used) for which the S/N was too low in the bluest part of the spectrum.

\cite{borsa_ox_nature} discussed the negligibility of line broadening due to the exposure time (i.e. ~4.7, 3.5, and 1.3 km/s on average for exposure times of 400, 300, and 111 s, respectively) for CARMENES data. In our case, combining all six nights, we obtained a mean exposure time of 433s, weighting the in-transit exposure times for the mean S/N of each of our 202 in-transit spectra. We therefore included the line broadening in our analysis with the NLTE models, as is discussed in Section \ref{sec:comp_models}.

\begin{table}[h!]
\caption{Stellar and planetary parameters of the KELT-9 system adopted in this work.}
\begin{tabular}{lll}
\hline
Parameter     & Value     & Reference               \\ \hline
$T_{_\mathrm{eff}}$ [K]      & 9600 $\pm$ 400     & \cite{borsa_ar}               \\
$M_{*}$ [M$_\odot$]     & 2.32  $\pm$ 0.16    & \cite{borsa_ar}        \\
$R_{*}$ [R$_\odot$]     & 2.418 $\pm$ 0.058    & \cite{borsa_ar}        \\
$\log_{10}(g)$     & 4.1 $\pm$ 0.3    & \cite{borsa_ar}       \\ 
$M_\text{p}$ [M$_\text{J}$]       & 2.88  $\pm$ 0.35    & \cite{borsa_ar}        \\
$R_\text{p}$ [R$_\text{J}$]       & 1.936 $\pm$ 0.047    & \cite{borsa_ar}        \\ 
$a$ [AU]        & 0.03368  $\pm$ 0.00078 & \cite{borsa_ar}        \\
$P$ [days]      & 1.48111871(16) & \cite{ivshina}       \\
$T_\text{c}$ [days]      & 2458415.362562(81)  & \cite{ivshina}       \\
$i$ [deg]       & 86.79  $\pm$ 0.25   & \cite{2017Natur.546..514G}       \\
$\lambda$ [deg]  & 85.78 $\pm$ 0.46     & \cite{borsa_ar}        \\
$K_\text{p}$ [km/s]    &    246    &       This work                  \\
$v_{_\text{sys}}$ [km/s]  &$-$17.74 $\pm$ 0.11   & \cite{hoeijmakers2019spectral} \\
$e$       & 0 (fixed)        & \cite{2017Natur.546..514G}       \\
$v$ $\mathrm{sin}$ $i_*$ [km/s] & 111.4  $\pm$ 1.2   & \cite{2017Natur.546..514G}      \\ \hline
\label{tab:system_params}
\end{tabular}
\end{table}

\section{Transmission spectra extraction} \label{sec:methods}
We employed the 3.7 version of the HARPS-N Data Reduction Software \citep[DRS;][]{2002A&A...388..632P} to do an initial reduction of the HARPS-N raw data, which includes a correction for the blaze function that takes care of both the instrumental blaze and the Earth atmospheric change thanks to the Atmospheric Dispersion Corrector. We operated with S1D spectra that were created by the DRS, starting from the S2D images. During the merging procedures, the DRS takes care of the overlaps between the 69 HARPS-N orders.
At this stage, the data contain the stellar signal, the planetary signal, and the telluric contamination, all given in the solar system barycentric reference frame. The wavelength information is given in air. For each transit, we calculated the transmission spectra of the single lines following the procedure described in \cite{wyttenbach2015spectrally}, and hence comparing out-of-transit and in-transit spectra.

\begin{figure}[ht]
   \centering
   \includegraphics[width=0.95\columnwidth]{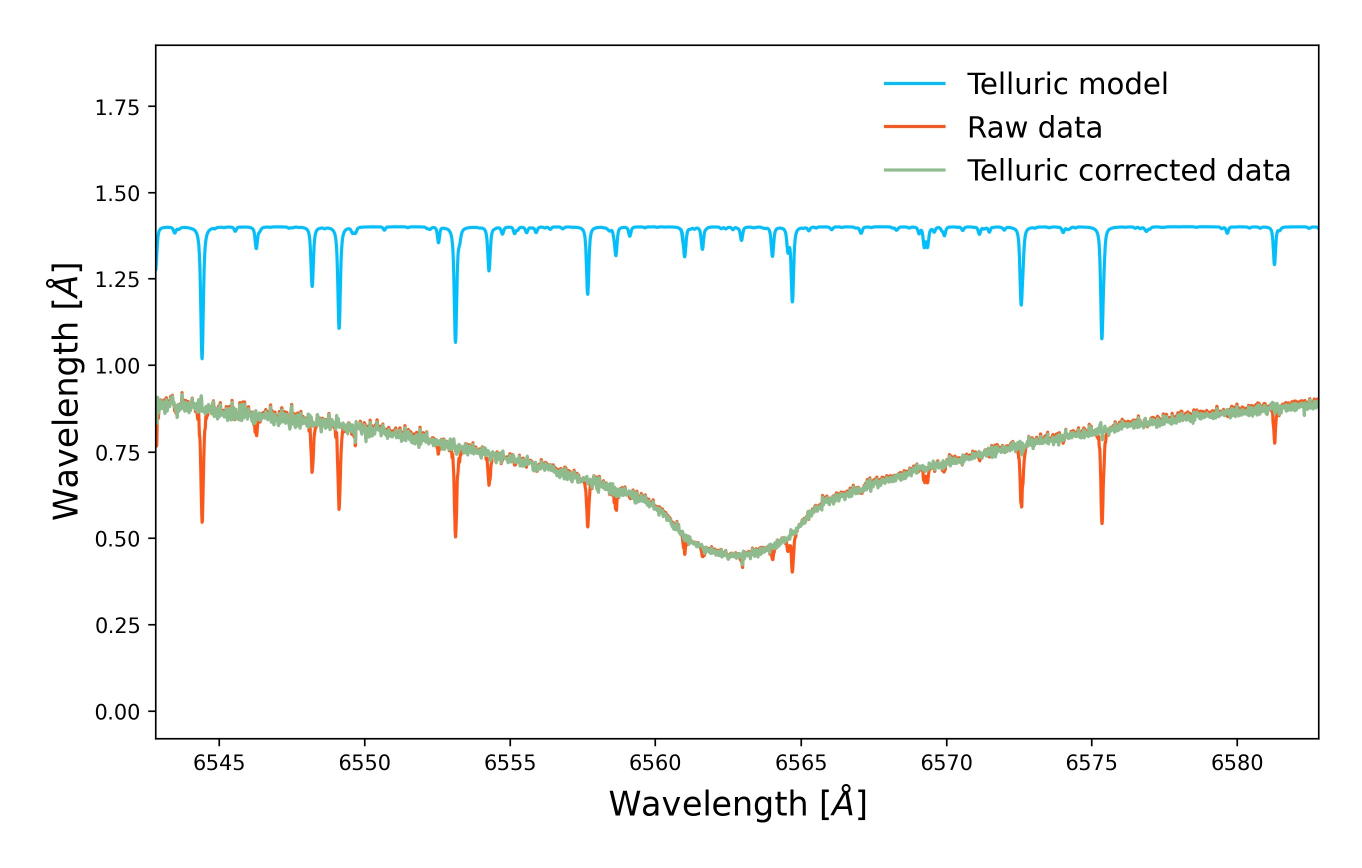}
      \caption{Example of telluric correction on a KELT-9 spectrum observed by HARPS-N on July 31 2017. After normalisation, the DRS-processed data (in red) was telluric-corrected with {\tt\string Molecfit} (green line) using the telluric model (in blue), here shifted upwards for clarity.}
    \label{fig:telluric_correction}
   \end{figure}

The first step of the analysis consists of correcting the observed spectra for telluric contamination. To do so, we employed {\tt\string Molecfit} \footnote{\url{http://www.eso.org/sci/software/pipelines/skytools/molecfit}} \citep{2015A&A...576A..77S, 2015A&A...576A..78K}, a specialised ESO tool designed to handle the correction of telluric atmospheric lines in astronomical spectra through the use of a line-by-line radiative transfer model. Although the software was specifically developed to correct data obtained with ESO instruments, in principle it is able to correct spectra from non-ESO ground-based spectrographs as well, and recently ESO released an additional experimental support for HARPS-N. 
Telluric correction is particularly important in the red part of the HARPS-N spectra, where the Earth atmosphere's H$_2$O and O$_2$ absorption features become relevant. In Fig.~\ref{fig:telluric_correction}, we show an example of telluric removal in the vicinity of the H$\alpha$ region. {\tt\string Molecfit} returned telluric-corrected spectra in the solar barycentric reference frame. Therefore, we shifted the spectra in the stellar reference frame by correcting for the star's systemic radial velocity, $v_\text{sys}$. The telluric-corrected spectra were then normalised to a common continuum level in a narrow wavelength range around each absorption line of interest. The normalisation around a small region should also remove all the possible wavelength gradients arising from changes in the overall transmission of the atmosphere and other slow changes in the spectrograph.

\begin{figure*}[h!]
    \includegraphics[width=0.66\columnwidth]{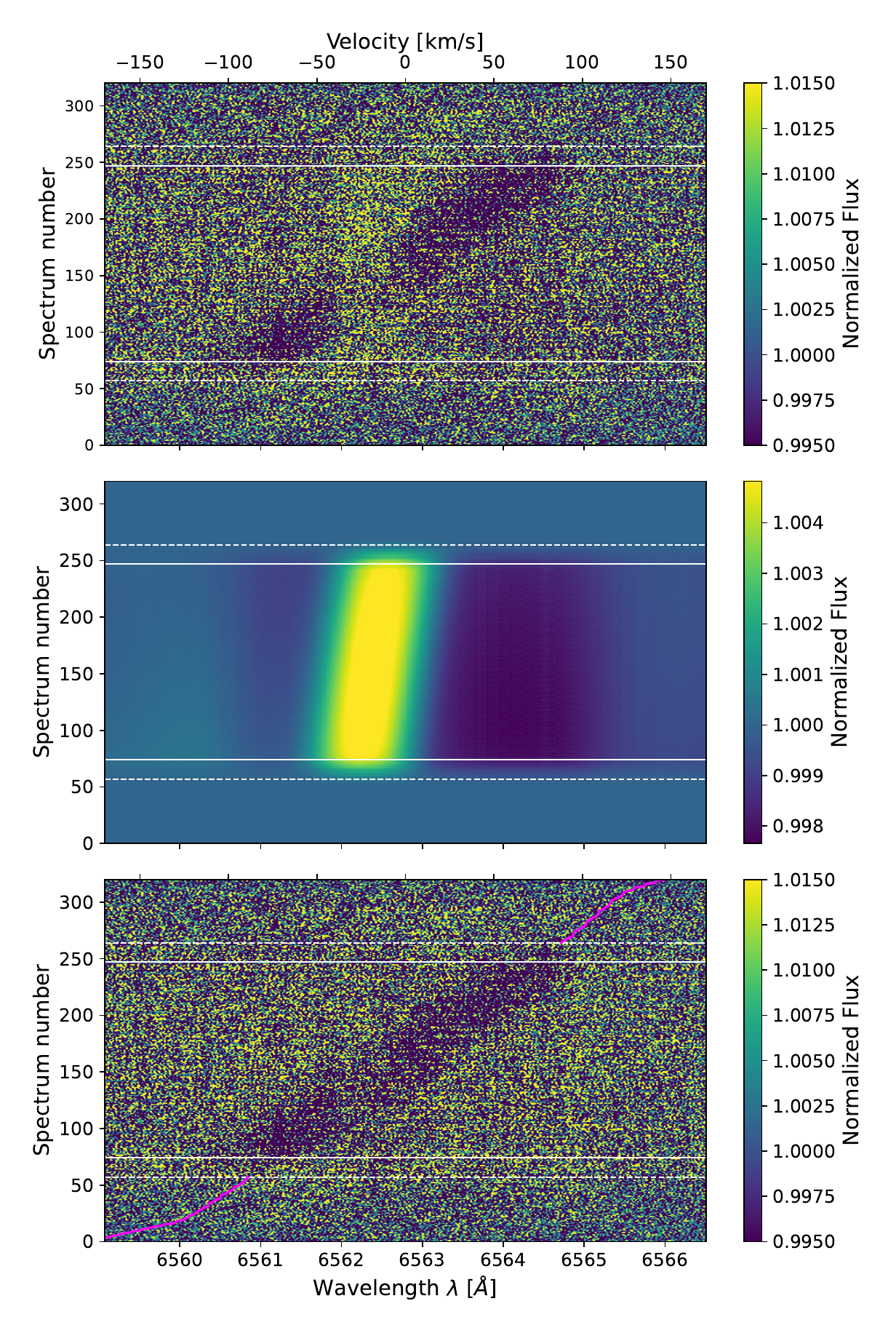}\quad
    \includegraphics[width=0.66\columnwidth]{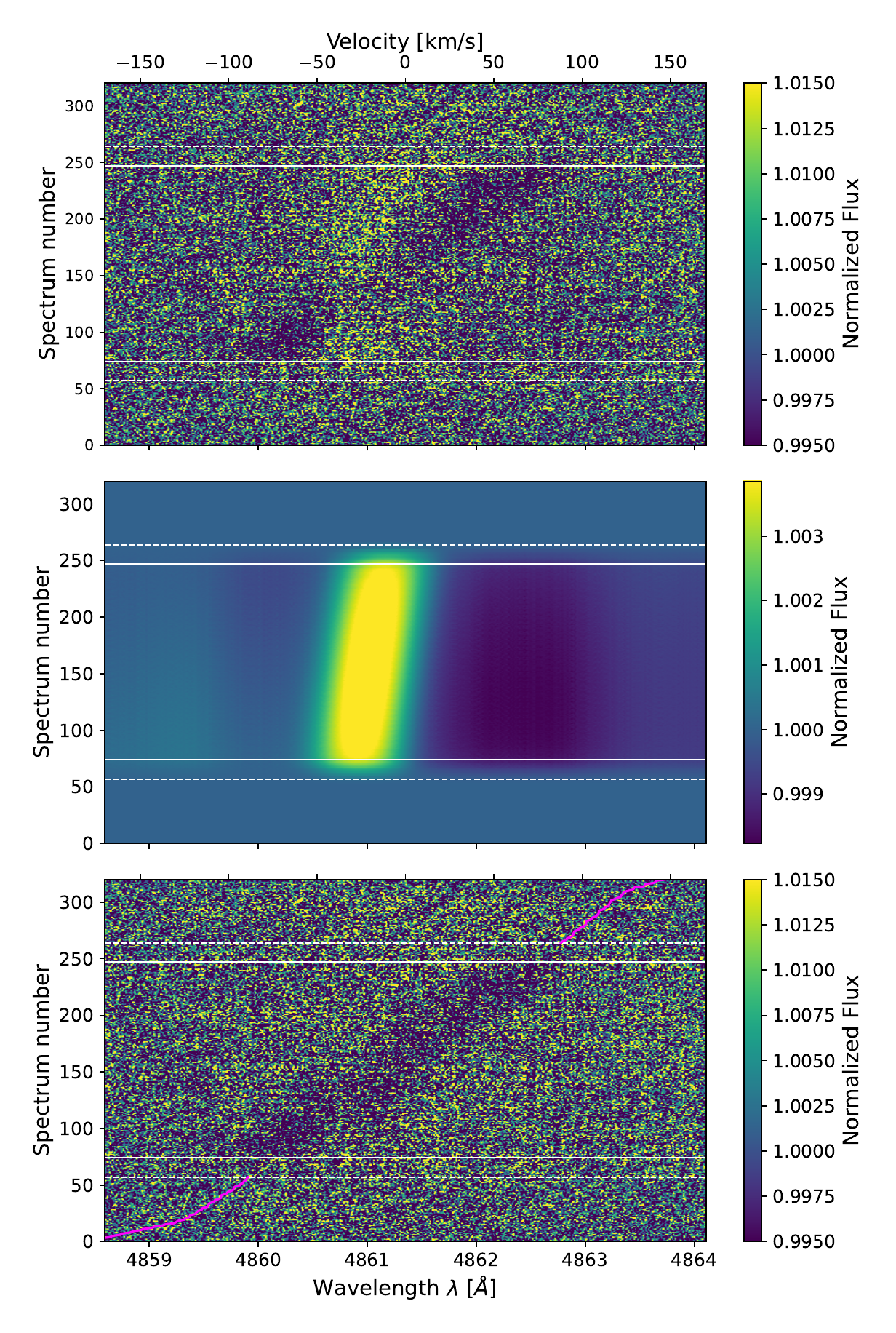}\quad
    \includegraphics[width=0.66\columnwidth]{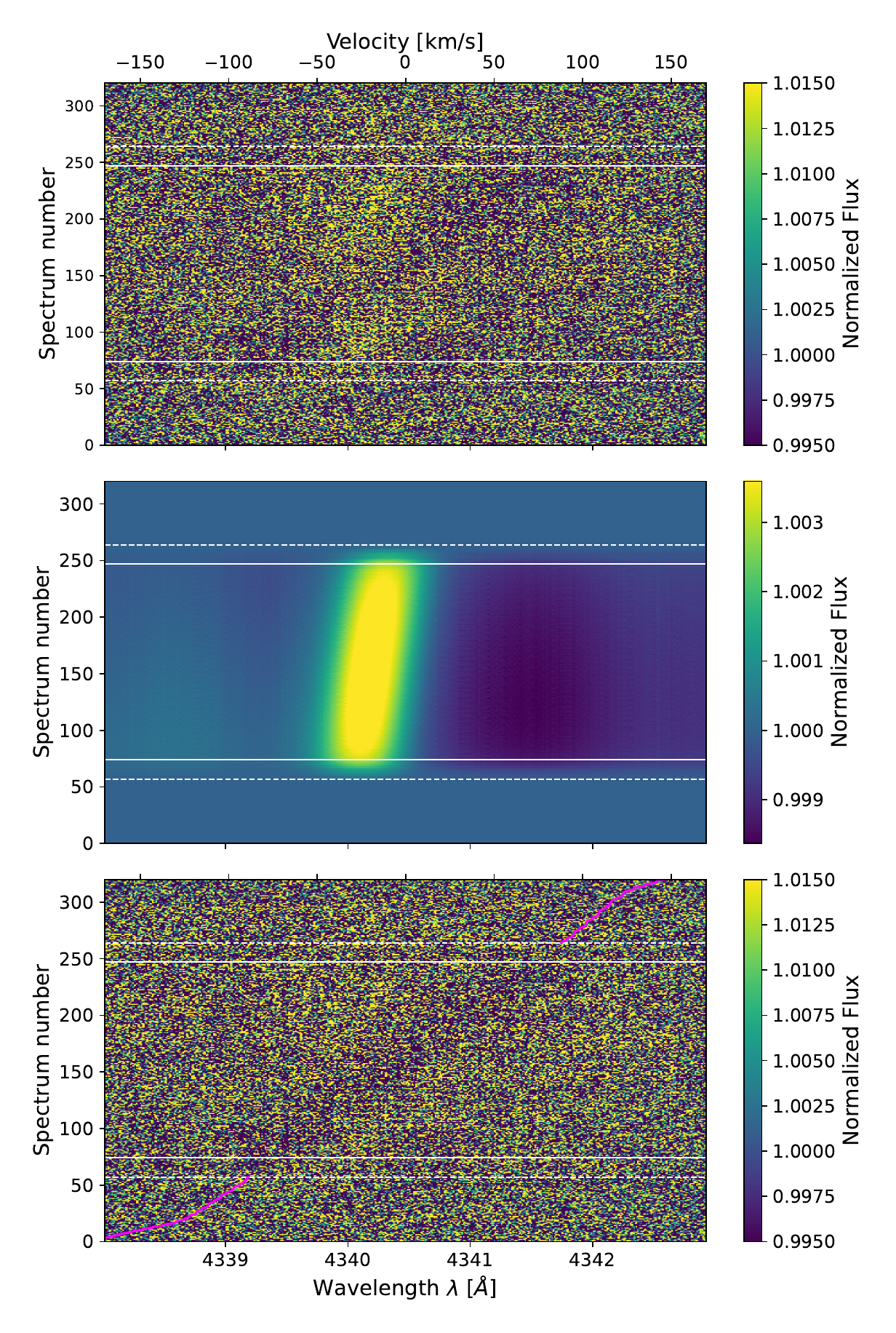}\quad
    \includegraphics[width=0.66\columnwidth]{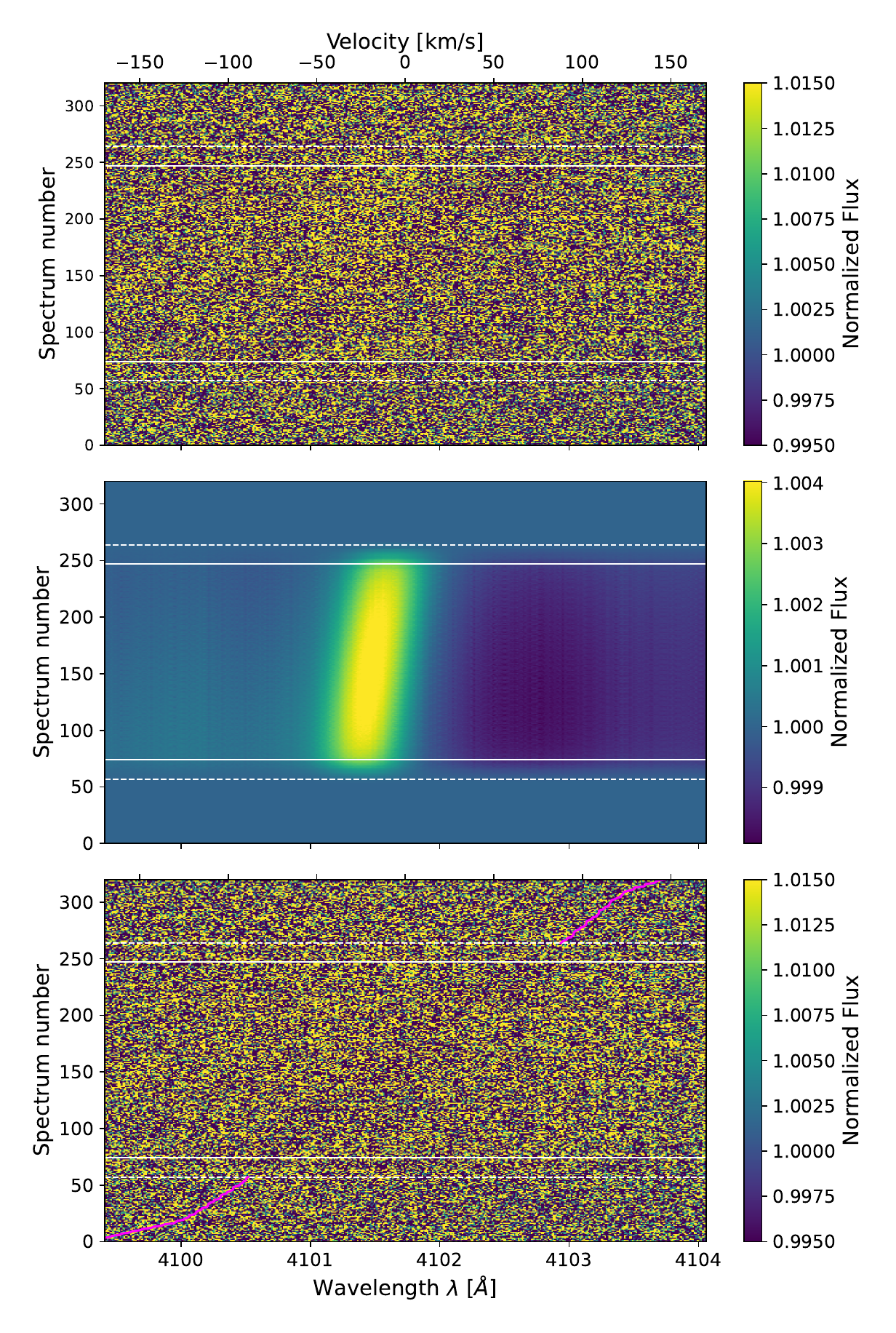}\quad
    \includegraphics[width=0.66\columnwidth]{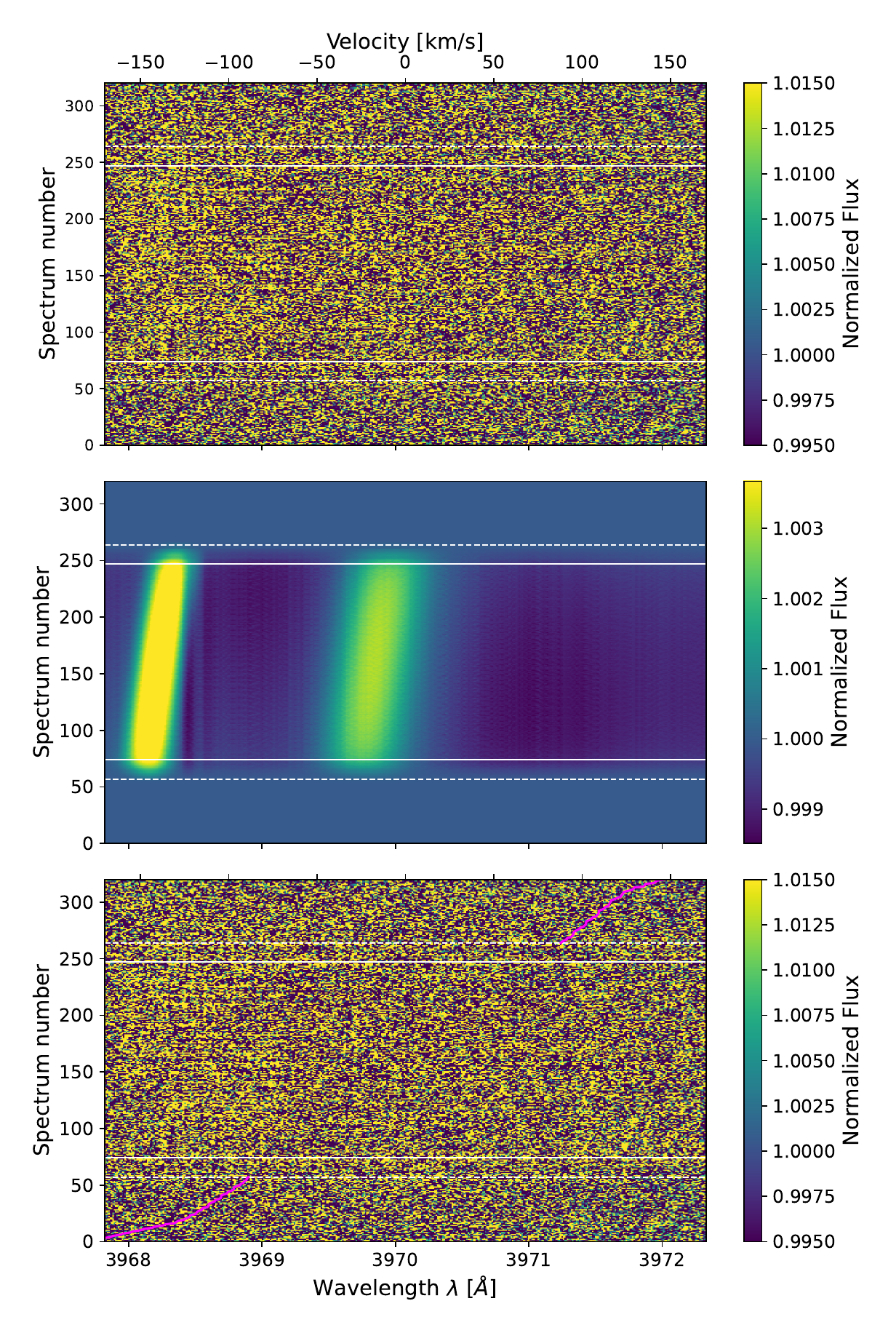}\quad
    \includegraphics[width=0.66\columnwidth]{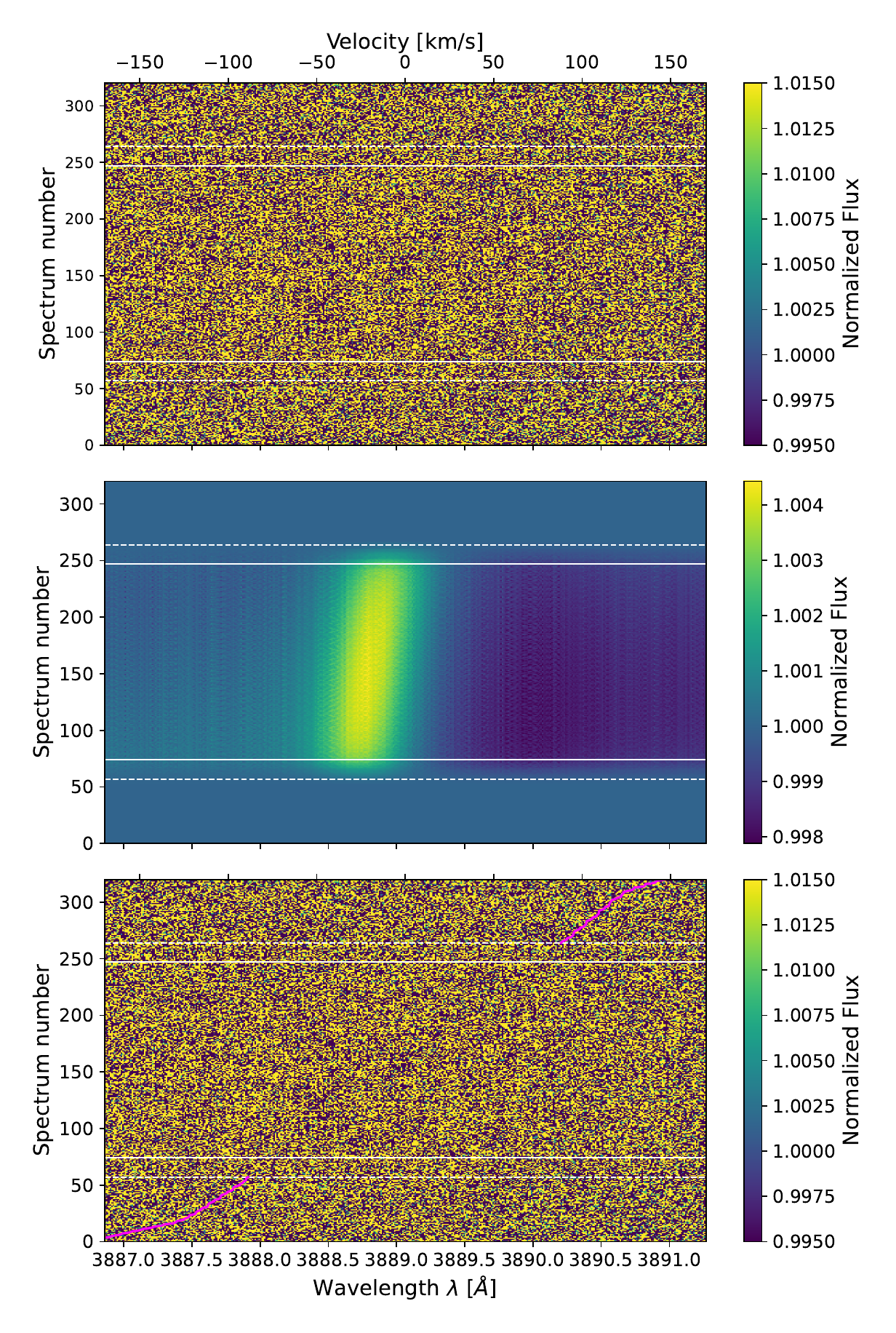}\quad

    \caption{ Tomographies in the stellar reference frame for the hydrogen Balmer lines.
    \textit{Top panels}: Raw tomographies  before the RME+CLV correction. Top row, left to right: H$\alpha$, H$\beta$, and H$\gamma$; Bottom row, left to right: H$\delta$, H$\epsilon$, and H$\zeta$. The planetary atmospheric absorption can be distinguished as a darker region following the expected planetary radial velocity profile, and the RME as a brighter feature. We also show the planetary radial velocity profile during out-of-transit with a magenta line. The contact points are represented with dashed and continuous white lines. \textit{Middle panels}: Model tomography computed in the manner explained in Section \ref{sec:methods}. \textit{Bottom panels}: Corrected tomography obtained by dividing the raw one by the model one.}
    \label{ha_toms}
\end{figure*}

For each night, we computed the transmission spectra in the stellar reference frame as the ratio between each spectrum and the weighted average out-of-transit spectrum (master-out). By stacking all spectra sorted by phase, we obtained a 2D map (referred to as tomography) such as the one shown in Fig.~\ref{ha_toms}. From this map, we can clearly see how the joint contribution of the Rossiter-McLaughlin effect (RME) \citep{rossiter1924detection, mclaughlin1924some} and the centre-to-limb variations (CLVs), which arise from the stellar rotation and the limb darkening, respectively, is a major contribution that needs to be modelled and removed. The RME and CLV compete with the planetary atmospheric signal to shape what we observe in the 2D map, such as Fig.~\ref{ha_toms}, according to the geometry of the transit.

We corrected for the RME+CLV effects by adopting an analogous method to that described by \cite{yan2017effect} and \cite{2017A&A...608A.135C}, and already applied in \cite{Guilluy2024}, which involves the use of stellar models. We used ATLAS9 stellar models \citep{kurucz1992model, 2005MSAIS...8...73K, 2014dapb.book...39K, 2017ascl.soft10017K} to compute the disc-integrated stellar model considering the system parameters listed in Table \ref{tab:system_params}. We computed the spectra for the case of a non-rotating star as well as a rotating star. We used {\tt\string PyLightCurve} \citep{tsiaras2016new} to calculate the planet's path during the transit to evaluate the part of the stellar disc obscured by the planet at each phase (hereafter, obscured region). We then built a 0.01 R$_*$ $\times$ 0.01 R$_*$ pixel size grid to approximate the stellar disc and for each phase we performed the following operations:
\begin{itemize}
    \item for each pixel belonging to the obscured region, we computed the intensity according to the limb darkening angle and the radial velocity shift with respect to the stellar rotation axis;
    \item we computed the spectrum of the obscured region using the ATLAS9 models, summing the spectrum of each pixel properly shifted by its own radial velocity.
\end{itemize}

Instead of using different model spectra at different limb darkening angles, $\mu$, and interpolating to compute the right intensity for each pixel, we used an analytical approach. We computed the limb darkening coefficients for our stellar model using the {\tt\string ExoTethys} \citep{morello2020exotethys} function {\tt\string Sail}, adopting a quadratic limb darkening law:
\begin{equation}
    \frac{I_{\lambda}(\mu)}{I_{\lambda}(1)} = 1 - c_{1,\lambda}(1-\mu) - c_{2,\lambda}(1-\mu)^2,
    \label{ld_eq}
\end{equation}
where: 
\begin{itemize}
    \item $\lambda$ indicates a specific spectral bin or passband;
    \item $\mu$ = cos $\theta$, $\theta$ being the angle between the line of sight and the normal to the stellar surface;
    \item $I_{\lambda}(\mu)$ is the stellar intensity profile and $I_{\lambda}(1)$ is the intensity at the centre of the disc;
    \item $c_{1,\lambda}$ and $c_{2,\lambda}$ are the limb darkening coefficients.
\end{itemize}
We note that the different choice of limb darkening law could be a possible cause of discrepancy with other analyses, since the RME+CLV modelling strongly relies on the part of the stellar disc occulted. However, since we do not have any spatially resolved spectrum of the star, it is not possible to state exactly which limb darkening law would better mimic the real observations.

The integrated flux, $F_{\lambda}$, as a function of the limb darkening angle, $\mu$, can be expressed as
\begin{equation}
    F_{\lambda} = 2\pi \int_0^1I_{\lambda}(\mu) \ \mu \ d\mu .
    \label{flux_eq}
\end{equation}

By combining Equations \ref{ld_eq} and \ref{flux_eq} and solving the integral analytically, we obtained $I_{\lambda}(1)$. At this point, knowing $I_{\lambda}(1)$ and the limb darkening coefficients, we were able to compute the flux for each $\mu$ (and hence for each pixel) using Eq. \ref{ld_eq}. The integrated flux, $F_{\lambda}$, corresponds to the non-rotating model spectrum. The rotational broadening is given by the sum of the different contributions of each pixel properly shifted by their own radial velocity. The observed master-out is broadened by the stellar rotation, and hence we used the rotationally broadened model as a comparison to the master-out to normalise the non-rotating model continuum. We then subtracted the modelled obscured region from the master-out spectra and divided all spectra by this quantity.
The average continuum for the RME+CLV models was set to one before applying the correction.
An example of the planet absorption for the Balmer series spectra in the stellar reference frame before and after the correction for RME and CLV is shown in Fig.~\ref{ha_toms}.
From the comparison between the top and bottom panels of Fig.~\ref{ha_toms}, we can see that, by following this framework, we were able to remove the RME+CLV contaminations, while not changing the planetary signal. This is particular easy to observe in a target such as KELT-9 b with its peculiar transit, but may be more challenging when the two effects overlap, such as in the case of HAT-P-67 b \citep{2024A&A...687A.143S}.
We have not considered the potential distortion of line profiles caused by gravity darkening. However, it is noteworthy that such distortions are rarely considered in the literature \citep{2021A&A...653A.104B}, and are anticipated to be negligible for this target, as is indicated by \cite{2022AJ....163..122C}, who state that the gravity darkening is expected to be less relevant with respect to the other effects considered (such as RME and CLV) with the present level of achievable precision. Furthermore, we did not include in our models broadening due to the exposure time, since the polar orbit of KELT-9 b causes the planet to obscure the same portion of the star in terms of velocity. In fact, during the 600s exposure the difference in the velocity occulted is below 1 km/s.
Once the spectra in the stellar reference frame had been corrected, we shifted all the in-transit spectra to the planet reference frame by correcting for its radial velocity profile in a circular orbit scenario,
\begin{equation}
    RV(\Phi) = K_\text{p} \ \sin{2\pi\Phi},
\end{equation}
where $\Phi$ is the orbital phase and K$_\text{p}$ is the radial velocity of the planet, computed using the parameters listed in Table \ref{tab:system_params}. The final transmission spectrum of each line was calculated as the error-weighted average of all the full in-transit spectra ($T_\text{2}$-$T_\text{3}$) in the planet reference frame. The uncertainties were originally computed as the square root of the DRS-corrected flux and then propagated during the entire process.

\section{Single-line analysis results} \label{sec:results}
For each analysed line, we extracted the transmission spectrum, as is described in Section \ref{sec:methods}, and then performed a Gaussian fit returning the absorption depth, the full width at half maximum (FWHM), and the velocity shift with respect to the line wavelength corresponding to a zero radial velocity. We used the Python package {\tt\string Scipy CurveFit}\footnote{\url{https://docs.scipy.org/doc/scipy/reference/generated/scipy.optimize.curve_fit.html}}, which employs non-linear least squares to fit a function, setting 100 km/s as the velocity range. For some lines, we adopted a smaller wavelength range (e.g. 80 km/s) due to the closeness with other strong lines.
We left the continuum offset as a free parameter and we used the retrieved values to normalise the spectra to one, and hence we do not report the continuum values in the best-fit tables.
Our definition of significance is based on the retrieved value of the Gaussian fit and it is defined as the ratio between the absorption depth value and its error.\\
Our procedure of modelling and removing of RME + CLV strongly depends on the size of the obscured region of the stellar disc, which, at the first order, we consider equal to the planetary radius. To include the atmospheric extension, we translated the line absorption depths, returned by the Gaussian fit, into planetary radii 
and used these values to repeat the RME + CLV removal with an increased size that accounts for both the planetary radius and the atmospheric extension. We repeated the loop, retrieving a new depth expressed in planetary radii, until the threshold convergence (0.001 $R_\text{p}$) was reached. 
Adopting this procedure, as opposed to \cite{yan2017effect}, we are including both the RME and the effective size of the planet occulting the stellar disc by using a wavelength-dependent line profile derived from the planet that accounts for the velocity change in the planetary atmosphere's signal.
Therefore, our correction, similar to other ones adopted for several instruments and targets \citep{2017A&A...608A.135C, 2019A&A...632A..69Y, borsa_rot, 2022A&A...662A.101S, kelt20}, manages to deal with both the planet's atmospheric lines and also the RM signal in the overlapping regions.
\subsection{Hydrogen Balmer series}
\label{sec:H}

We focussed on the hydrogen Balmer series by analysing all the lines observable in the HARPS-N wavelength range, spanning from H$\alpha$ to H$\zeta$.
In Fig.~\ref{balmer_sp}, we show the transmission spectra around each Balmer line.
We significantly detect all the lines observable in the available wavelength range with the first significant detection of H$\epsilon$ and H$\zeta$. The parameters of the Gaussian best fit are listed in Table \ref{balmer_tab}.\\

\subsubsection{List of Balmer lines detected}
- H$\alpha$: We detect H$\alpha$ with a significance of $\sim 60 \sigma$. H$\alpha$ has already been detected in this target by \cite{yan2018extended} and \cite{turner2020detection} with CARMENES, and by \cite{cauley2019atmospheric} using PEPSI. The Balmer series has been analysed by \cite{wyttenbach2020mass} using Night 1 and Night 3 data, detecting H$\alpha$, H$\beta$, H$\gamma$, and H$\delta$, but not H$\epsilon$ (tentative detection) and H$\zeta$ (non detection). As is shown in Fig.~A.2, our results are in good agreement (<3$\sigma$) with \cite{turner2020detection} and \cite{wyttenbach2020mass}, but in disagreement (>5$\sigma$) with \cite{cauley2019atmospheric} and \cite{yan2018extended}.\\ 
- H$\beta$: We detect H$\beta$ with a significance of $\sim 31 \sigma$ and, as for H$\alpha$, our result agrees (2$\sigma$) with \cite{wyttenbach2020mass}, but it is at odds (7$\sigma$) with \cite{cauley2019atmospheric}.\\
- H$\gamma$ and H$\delta$: These lines have been detected in the atmospheres of just a few exoplanets \citep[HD189733\,b, KELT-9\,b, and KELT-20\,b/MASCARA-2\,b;][]{cauleyhd, cauley2019atmospheric, 2019A&A...628A...9C}, with \cite{wyttenbach2020mass} being the only study to have detected both lines in the atmosphere of KELT-9\,b. We detect H$\gamma$ and H$\delta$ with a significance of 25$\sigma$ and 12.2$\sigma$, respectively, further supporting \cite{wyttenbach2020mass} detection, especially for H$\delta$.\\
- H$\epsilon$: We significantly detect H$\epsilon$ for the first time in the atmosphere of an exoplanet with a significance of 6.8$\sigma$. H$\epsilon$, already marginally detected by \cite{wyttenbach2020mass}, falls in a region in which two other strong lines are present; namely, Ca {\sc ii} H at 3968.47 $\AA$ and Fe {\sc i} at 3969.25 $\AA$. The Ca {\sc ii} H line is discussed in Section \ref{sec:ca}; it is more than 100 km/s away from the core of H$\epsilon$, and hence it does not affect our detection. The Fe {\sc i} line lies between the Ca {\sc ii} H lines and H$\epsilon$, and we find that is shifted by 8.67 km/s, similar to the other Fe lines (see Section \ref{sec:fe}). The region including Ca {\sc ii} H, Fe {\sc i}, and H$\epsilon$ is shown in Fig.~A.3 with the corresponding Gaussian fits.\\
- H$\zeta$: We detect H$\zeta$ for the first time in the atmosphere of an exoplanet, with a significance of 5.7$\sigma$. We find that it is deeper than the upper limit measured by \cite{wyttenbach2020mass}, who expected the absorption to be the smallest among the Balmer lines in the HARPS-N range. Instead, we find H$\zeta$ to be the second deepest line after H$\alpha$, though with a very large uncertainty (almost three to ten times larger than for the other lines). The discrepancy is most likely due to the low S/N in the bluest region covered by HARPS-N that forced us to discard eight spectra because of their low S/N. The low signal significantly affects both the uncertainties and the normalisation process, because of the difficulty of finding a pseudo-continuum region around the line core, which may have led to overestimating the line depth.
We also investigated the possibility that the larger depth is due to the presence of other lines in the region; namely, Fe {\sc i} at 3888.51 $\AA$ and Ca {\sc i} at 3889.10 $\AA$ lines. We analysed these lines adopting the procedure described in Section \ref{sec:methods} and found that the depth and FWHM are similar to those obtained analysing H$\zeta$, while the velocity shifts in both cases correspond to the exact position of H$\zeta$. Furthermore, the analysis of the Ca {\sc i} and Fe {\sc i} lines presented in Sections \ref{sec:ca} and \ref{sec:fe} reveals that the detected lines of those species are shallower compared to those obtained for Ca {\sc ii} at 3889.10 $\AA$ and Fe {\sc i} at 3888.51 $\AA$.

\subsubsection{Comparison with previous works and night-to-night variability}

A detailed discussion about the differences in the H$\alpha$ and H$\beta$ transmission spectra observed by \cite{cauley2019atmospheric}, \cite{yan2018extended}, \cite{turner2020detection} and \cite{wyttenbach2020mass} is given by \cite{fossati_comp}. Their conclusions stress that the variations could stem from distinct instruments and resolving powers, but more likely they arise from the slightly differing methodologies employed to extract the planetary signal. 
The major contributions to the discrepancies seem to be the normalisation, the removal of the RME+CLV effects, and the systemic velocity adopted in the different studies, with the latter being responsible for the velocity shifts \citep[see][for more details]{fossati_comp}.
We can extend the conclusions that \cite{fossati_comp} drew for the other works to our results, which fit well in the scenario, considering the slightly different approaches used in the analyses. In Fig.~A.2, we show that the H$\alpha$ results are in good agreement (<3$\sigma$) with those of \cite{turner2020detection} and \cite{wyttenbach2020mass}, but at odds with those of \cite{cauley2019atmospheric} and \cite{yan2018extended}, while H$\beta$ is in good agreement with \cite{wyttenbach2020mass}, but not with \cite{cauley2019atmospheric}.

Another plausible explanation for the discrepancy may be the intrinsic presence of in-transit variations along a single transit, as was pointed out by \cite{cauley2019atmospheric}.
Thanks to our sample of six nights, we were able to explore the presence of variability using our method by analysing H$\alpha$, H$\beta$, and H$\gamma$ individually for each night. Fig.~A.2 and Table~A.1 show that the individual nights' best-fit results scatter around the value obtained by analysing all the nights together. None of the parameters seem to show a trend that depends on the different exposure time used in the first three nights (600s) and in the last three (300s).
Focussing on the H$\alpha$ absorption depths, we see an intrinsic variability of the order of $\sim$30\% in our dataset, with depths spanning over a range (0.78  $\pm$ 0.05 - 1.18  $\pm$ 0.03) that includes all the results from other works. Since our datasets have two common transit observations (July 31 2017 and July 20 2018), we deepened the comparison with \cite{wyttenbach2020mass}.
Fig.~A.5 shows that: i) our results are generally in good agreement (<3$\sigma$) with those of \cite{wyttenbach2020mass}, except for the velocity shift of H$\epsilon$, which has been previously only tentatively detected; and ii) the best-fit parameters considering two nights are closer to the ones that we obtained with six nights than the results obtained analysing the single nights separately, hinting that increasing the global S/N plays a major role in the final results.

We conclude that the discrepancies may generally arise from the joint contribution of several factors, such as different instrumentation and datasets, or they originate in the slightly different methods used to extract the planetary signal, as has already been discussed by \cite{fossati_comp}, or from the different $K_\text{p}$ used.
However, the analysis of six nights using the same instrument and the same framework hints at an intrinsic variability that covers the range of the different values retrieved in the different works. The possible causes of this variability are still unclear, but one cannot exclude that it originates in the planetary atmosphere. A similar intrinsic variability has already been noted by \cite{kelt20} in analysing HARPS-N data of UHJ KELT-20 b, despite, in their case, the amplitude of the variation not being significant.

\begin{figure*}[h!]
    \includegraphics[width=\columnwidth]{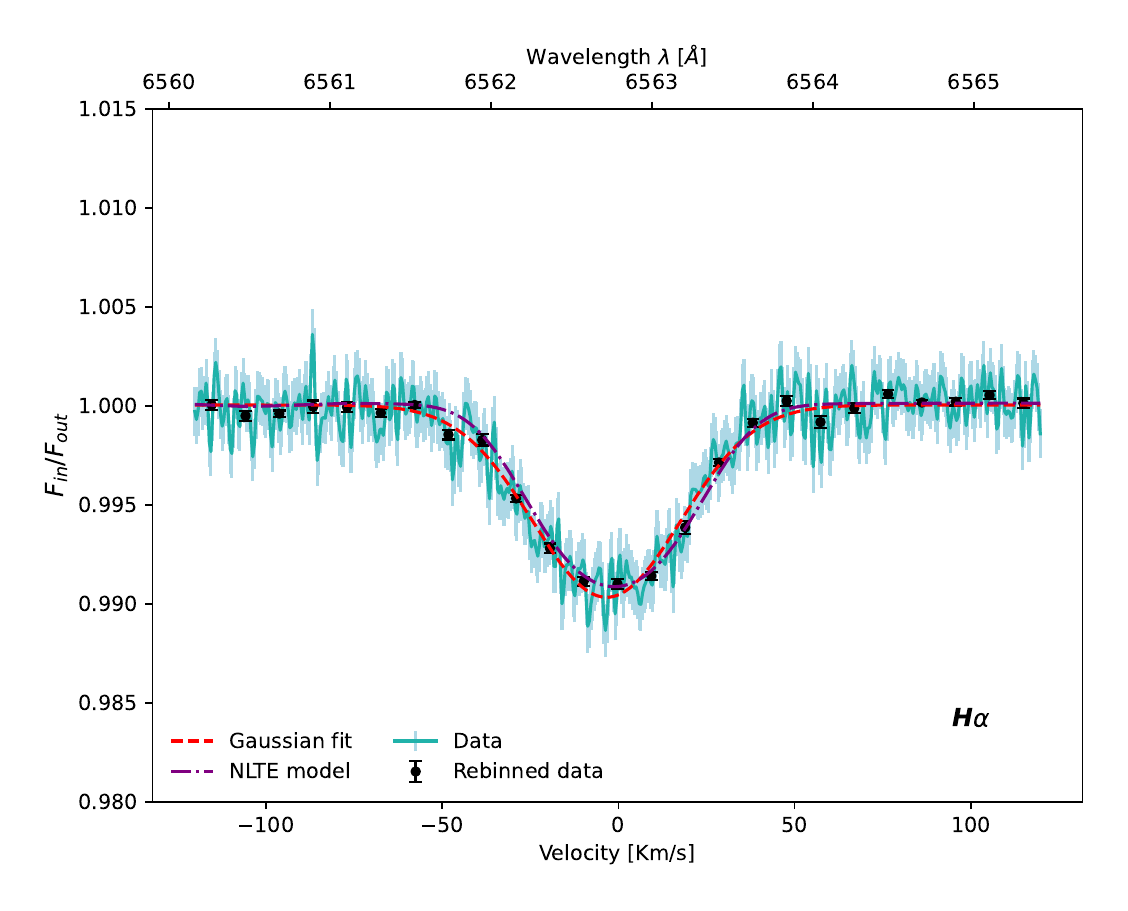}\quad
    \includegraphics[width=\columnwidth]{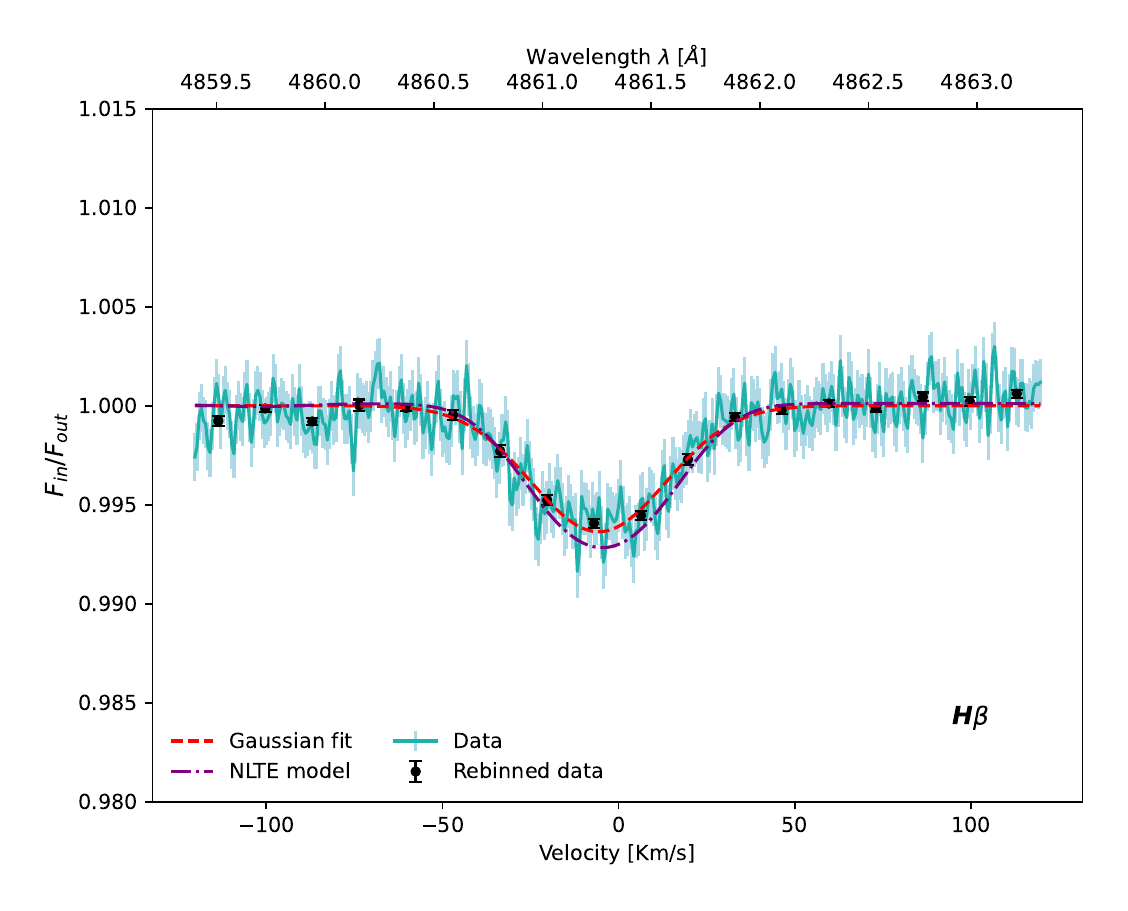}\quad
    \includegraphics[width=\columnwidth]{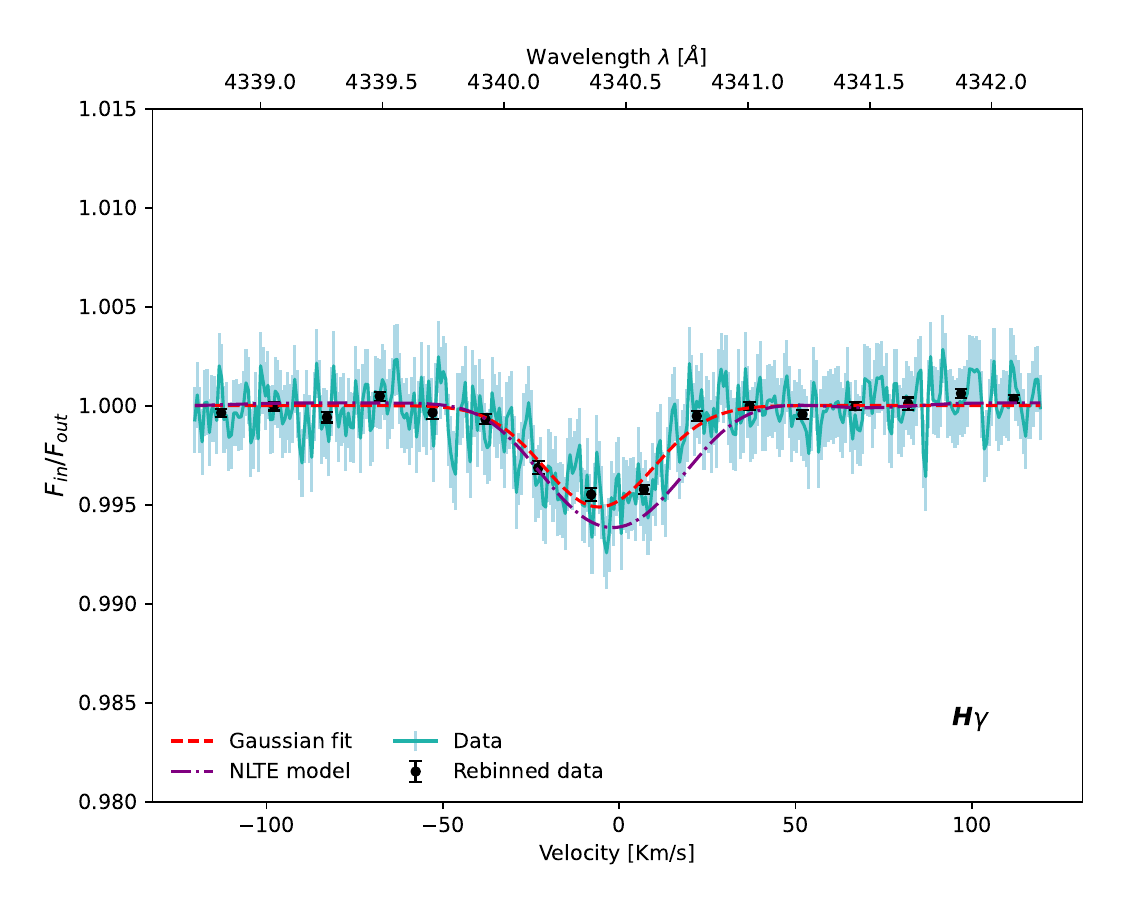}\quad
    \includegraphics[width=\columnwidth]{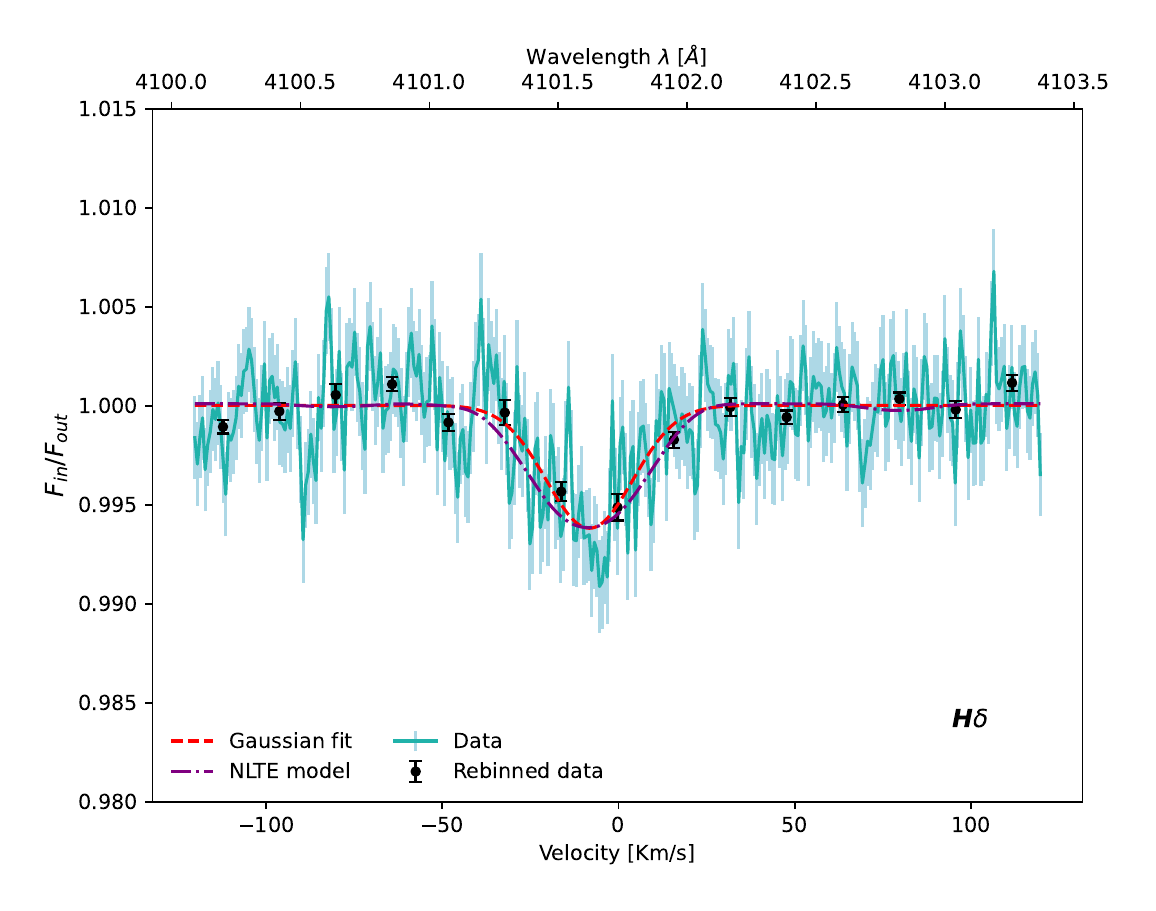}\quad
    \includegraphics[width=\columnwidth]{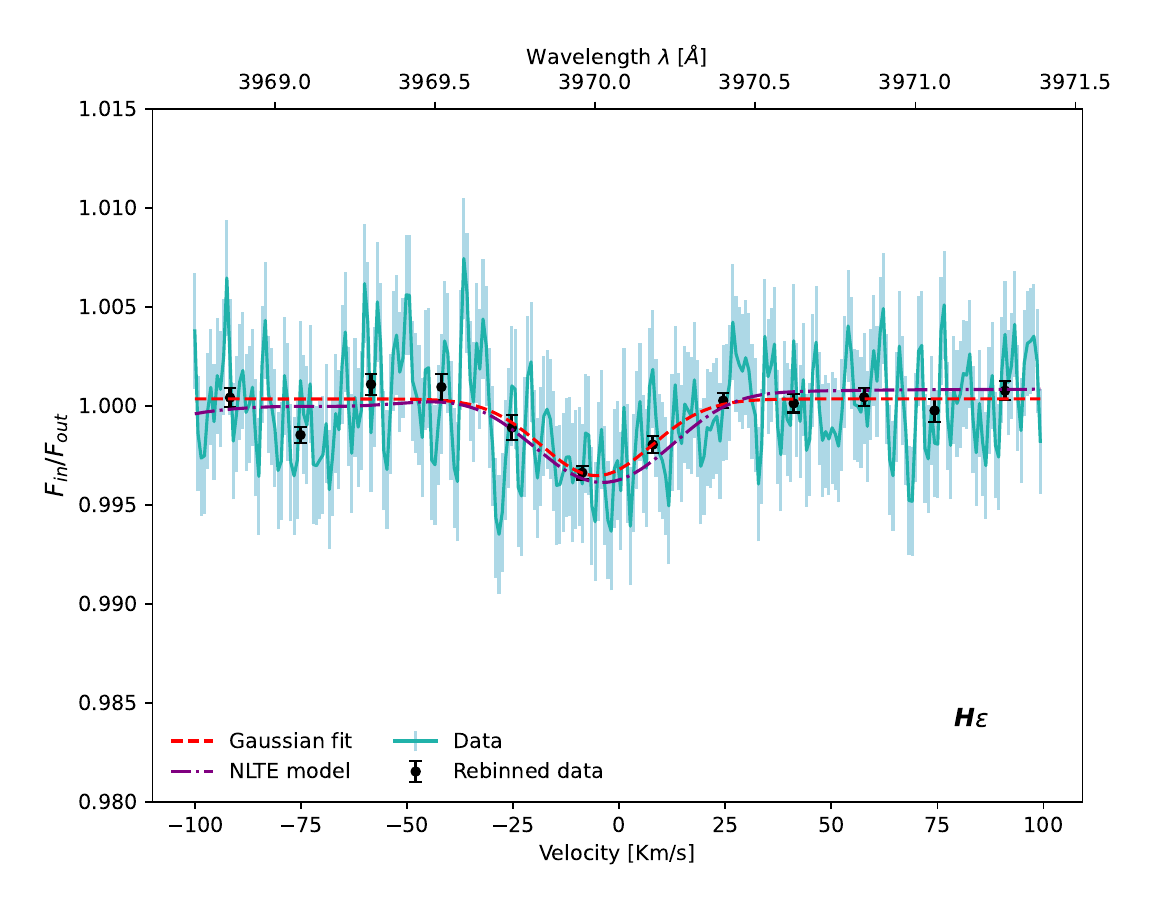}\quad
    \includegraphics[width=\columnwidth]{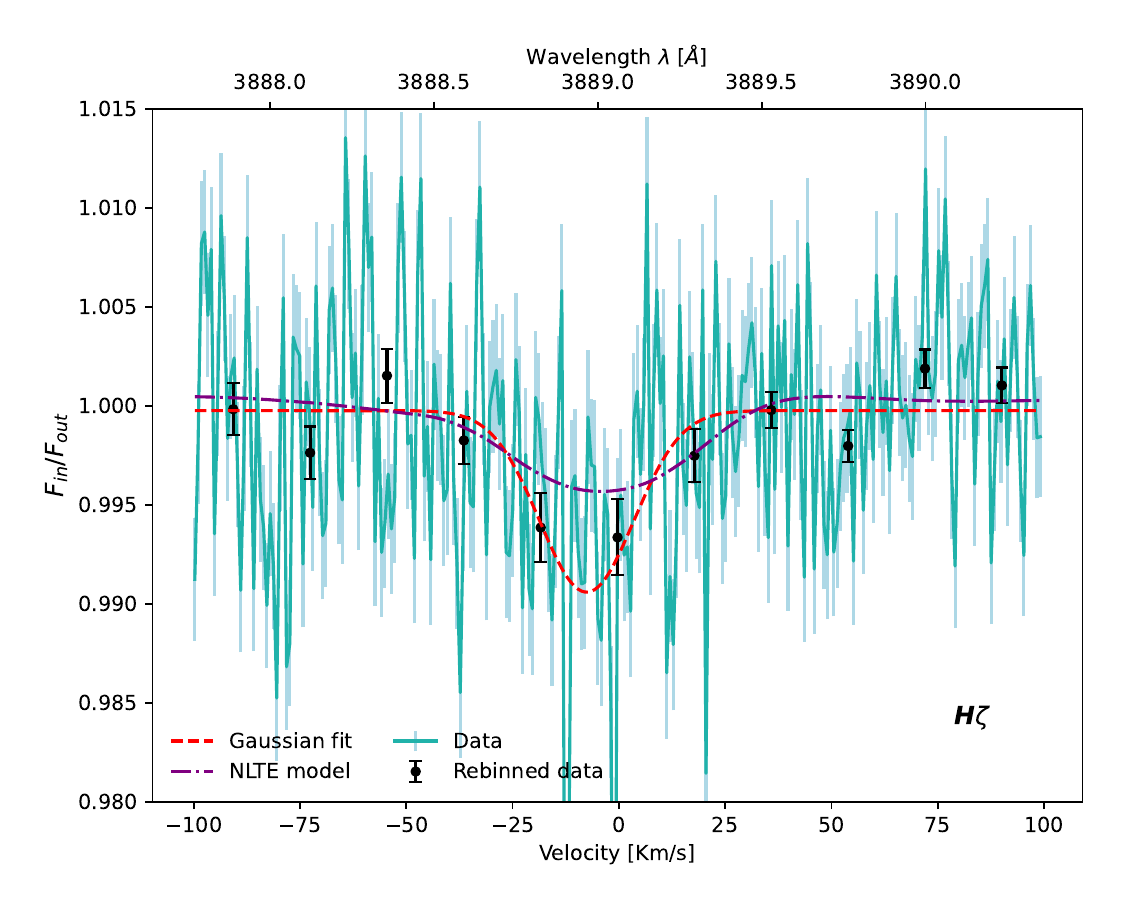}
    
    \caption{Balmer series transmission spectra: all the figures have the same range on the y axis to underline the differences between the single lines. The dashed red lines represent the Gaussian best fit, the results of which are listed in Table \ref{balmer_tab}, while the black dots are for a 20x binning. H$\zeta$ presents a wider scatter compared to the other lines due to the low S/N in the bluest part of the HARPS-N range. The purple lines show the NLTE broadened models discussed in Section \ref{sec:comp_models}. The small gradient of the NLTE model for H$\epsilon$ is due to the vicinity with the Ca H line at 3968.47 $\AA$.}
    \label{balmer_sp}
\end{figure*}

\begin{table*}[]
\caption{Summary of the results obtained from the Gaussian fits to the detected H {\sc i} Balmer lines.}
\centering
\begin{tabular}{l|lllll|lll}
\hline
&&& Gaussian fit &&&& NLTE model \\
\hline
Line  \tablefootmark{a}   & Depth [\%] &  Rp   & FWHM [km/s]    & v [km/s] & $\chi^2_\text{red}$ & $\nu_{\mathrm mic}$ [km/s] & $\nu_{\mathrm mac}$ [km/s] & $\chi^2_\text{red}$
\\
\hline
H$\alpha$ & 0.97 $\pm$ 0.02 & 1.56 $\pm$ 0.01 & 49.25 $\pm$ 0.95 & -3.34 $\pm$ 0.36 & 1.37 & 2.0 $\pm$ 1.5 & 21 $\pm$ 3 & 0.66\  \\
H$\beta$ & 0.64 $\pm$ 0.02 & 1.39 $\pm$ 0.01 & 45.0 $\pm$ 1.58 & -5.67 $\pm$ 0.61 & 1.08 & 1.0 $\pm$ 0.8 & 24 $\pm$ 1 & 0.71\  \\
H$\gamma$ & 0.51 $\pm$ 0.03 & 1.33 $\pm$ 0.01 & 36.63 $\pm$ 2.17 & -5.74 $\pm$ 0.86 & 0.66 & 3.0 $\pm$ 0.5 & 24 $\pm$ 2 & 0.6\  \\
H$\delta$ & 0.62 $\pm$ 0.05 & 1.38 $\pm$ 0.03 & 28.97 $\pm$ 2.79 & -8.24 $\pm$ 1.13 & 1.03 & 1.0 $\pm$ 1.2 & 10 $\pm$ 12 & 0.9\  \\
H$\epsilon$ & 0.39 $\pm$ 0.06 & 1.25 $\pm$ 0.03 & 30.65 $\pm$ 5.35 & -5.36 $\pm$ 2.11 & 0.7 & 1.0 $\pm$ 1.7 & 19 $\pm$ 16 & 0.83\  \\
H$\zeta$ & 0.92 $\pm$ 0.16 & 1.53 $\pm$ 0.08 & 26.1 $\pm$ 5.4 & -7.56 $\pm$ 2.16 & 3.6 & 14.0 $\pm$ 4.8 & 1 $\pm$ 14 & 3.75\  \\

\hline
\label{balmer_tab}
\end{tabular}
\tablefoot{ We report the Depth, FWHM, and v columns representing the Gaussian best-fit parameters, while $R_{\mathrm p}$ is the transit depth translated into planetary radii. $\nu_{\mathrm mic}$ and $\nu_{\mathrm mac}$ are the micro- and macro-turbulence velocity, respectively, for the best-fitting NLTE model. For both analyses, we also report the reduced chi-squared, $\chi^2_\text{red}$. \tablefoottext{a}{For each line, we show the wavelength in $\AA$.}}
\end{table*}

\subsection{Metal lines} 
We analysed several metal species by employing the same method used for the Balmer series.
We used the NIST Atomic Spectra Database Lines Data\footnote{https://physics.nist.gov/PhysRefData/ASD/lines\_form.html} to aid identifying the strongest lines in the HARPS-N range for each species, focussing on both neutral and ionised species.
We do not detect any He (listed here despite not being a metal), Li, Be, K, Sr, Y, Ni, and Ba lines. We stress that these species have not been detected on KELT-9\,b in previous studies \citep{hoeijmakers2018atomic, yan2018extended, cauley2019atmospheric, hoeijmakers2019spectral, borsato2023mantis, ridden2023high} neither via single-line analysis nor by using cross-correlation with templates. Instead, we detect Na, Mg, Ca, Sc, Ti, V, Cr, and Fe lines with a significance greater than 3$\sigma$. 
In the following sections, we list our results for each species in order of ascending atomic number. The best-fit parameters are listed in Table A.2.

\subsubsection{Sodium}
\label{sec:na}
The neutral sodium doublet lines, named Na D1 and Na D2 in this study, are among the most studied and detected lines in single-line analyses due to their capability to probe the upper layers of the atmosphere. The sodium doublet has already been detected in KELT-9\,b by \cite{2022MNRAS.514.5192L} using HARPS-N data collected during Night 1 and Night 3. Their results are in excellent agreement with ours, since they retrieve a mean depth of 0.16 $\pm$ 0.03 \%, while we retrieve 0.15 $\pm$ 0.02 \% and 0.16 $\pm$ 0.01 \% for Na D1 and Na D2, respectively. The mean velocity shift they retrieve is -4.1 $\pm$ 2.9, while ours is -4.59 $\pm$ 0.84 and -6.68 $\pm$ 1.05 for Na D1 and Na D2, respectively. The sodium doublet will also be analysed, using the same dataset employed in this work, in a dedicated study \citep{Siciliainprep} that aims to perform a population study on several GAPS targets.

\subsubsection{Magnesium}
\label{sec:mg}
We identify a significant absorption on each of the neutral Magnesium b triplet lines ($\sim$ 5167-5183 $\AA$), as has previously been reported by \cite{cauley2019atmospheric}, plus the Mg {\sc i} line at 5528.40 $\AA$. Our Mg {\sc i} triplet line depths are more than 3$\sigma$ away from the results reported by \cite{cauley2019atmospheric}, except for the Mg {\sc i} 5167.23 $\AA$ line, which is consistent within 1$\sigma$. Instead, our FWHM results are more in agreement (1$\sigma$ difference) with \cite{cauley2019atmospheric}, the first line at 5167.3216 $\AA$ being 2$\sigma$ away. The line shifts have remarkably large error bars, similar to \cite{cauley2019atmospheric}. The discrepancies in the fitted values between our study and \cite{cauley2019atmospheric} are probably due to the latter employing lower-resolution spectra from PEPSI ($\mathpzc{R}$$\sim$50,000) compared to HARPS-N. The Mg {\sc i} lines at 5167.23 and 5172.70 $\AA$ are located in the vicinity of the Fe {\sc ii} 5169.0282 $\AA$ line, as is highlighted by \cite{hoeijmakers2019spectral} in their Fig.~8. To avoid any potential contamination from this iron line, we restricted the velocity range around the centre of the reddest line to $\pm$ 80 km/s. This excludes the Fe {\sc ii} line, which does not appear in either transmission spectrum (see Fig.~A.7). For the first time, we detect an additional Mg {\sc i} line at 5528.40 $\AA$ with a 4.5$\sigma$ significance.

\subsubsection{Calcium}
\label{sec:ca}
We extracted transmission spectra for five Ca {\sc i} and two Ca {\sc ii} lines. The best Gaussian fit parameters are listed in Table A.2, while the transmission spectra are shown in Fig.~A.8. We detect for the first time five individual lines of neutral calcium with a significance larger than 3$\sigma$ and with absorption depths of 0.04--0.08\% that translate into planetary radii of 1.03--1.06 $R_\text{p}$. The region around these lines is not affected by other species observed in the atmosphere of KELT-9\,b, supporting their identification. This is the first detection of individual Ca {\sc i} lines in the atmosphere of KELT-9\,b, though Ca {\sc i} has been detected in the atmosphere of KELT-9\,b by \cite{borsato2023mantis} through using the cross-correlation technique on HARPS-N Night 1 and Night 3 data along with two nights obtained with CARMENES.

We also detect the Ca {\sc ii} H$\&$K doublet, finding an absorption depth of 0.67$\pm$0.05$\%$ and 0.54$\pm$0.06$\%$, which translates into planetary radii of 1.42 $R_\text{p}$ and 1.33 $R_\text{p}$, respectively. Ca {\sc ii} has been detected by \cite{borsato2023mantis} and \cite{2020ApJ...888L..13T} as well. The Ca {\sc ii} H$\&$K lines have been identified by \cite{yan2019ionized} using both cross-correlation and single-line analysis. \cite{yan2019ionized} reports a combined H$\&$K line depth of 0.78$\pm$0.04 $\%$, corresponding to 1.47$\pm$0.02 $R_\text{p}$, in agreement with our measurement of the Ca {\sc ii} H line. We already discussed the fact that the Ca {\sc ii} H line is blended with the H$\epsilon$ and Fe {\sc i} lines, as is shown in Fig.~A.3. However, by choosing a narrower range around the line, we were able to avoid the contamination and perform a Gaussian fit.

\subsubsection{Scandium}
\label{sec:sc}
Singly ionised scandium displays a few strong lines in the HARPS-N range, among which we could identify five (see Fig.~A.9 and Table A.2) ranging between 1.06 $R_\text{p}$ and 1.15 $R_\text{p}$. Despite ionised scandium also being identified through cross-correlation by both \cite{hoeijmakers2019spectral} and \cite{borsato2023mantis} with a high significance (>5$\sigma$ and >10$\sigma$, respectively), this is the first time that Sc II individual lines have been observed. We also searched for neutral scandium signatures, but failed to identify any of them, which supports the non-detection by \cite{hoeijmakers2019spectral} obtained via cross-correlation.

\subsubsection{Titanium}
\label{sec:ti}
\begin{figure*}
\sidecaption
\includegraphics[width=12cm]{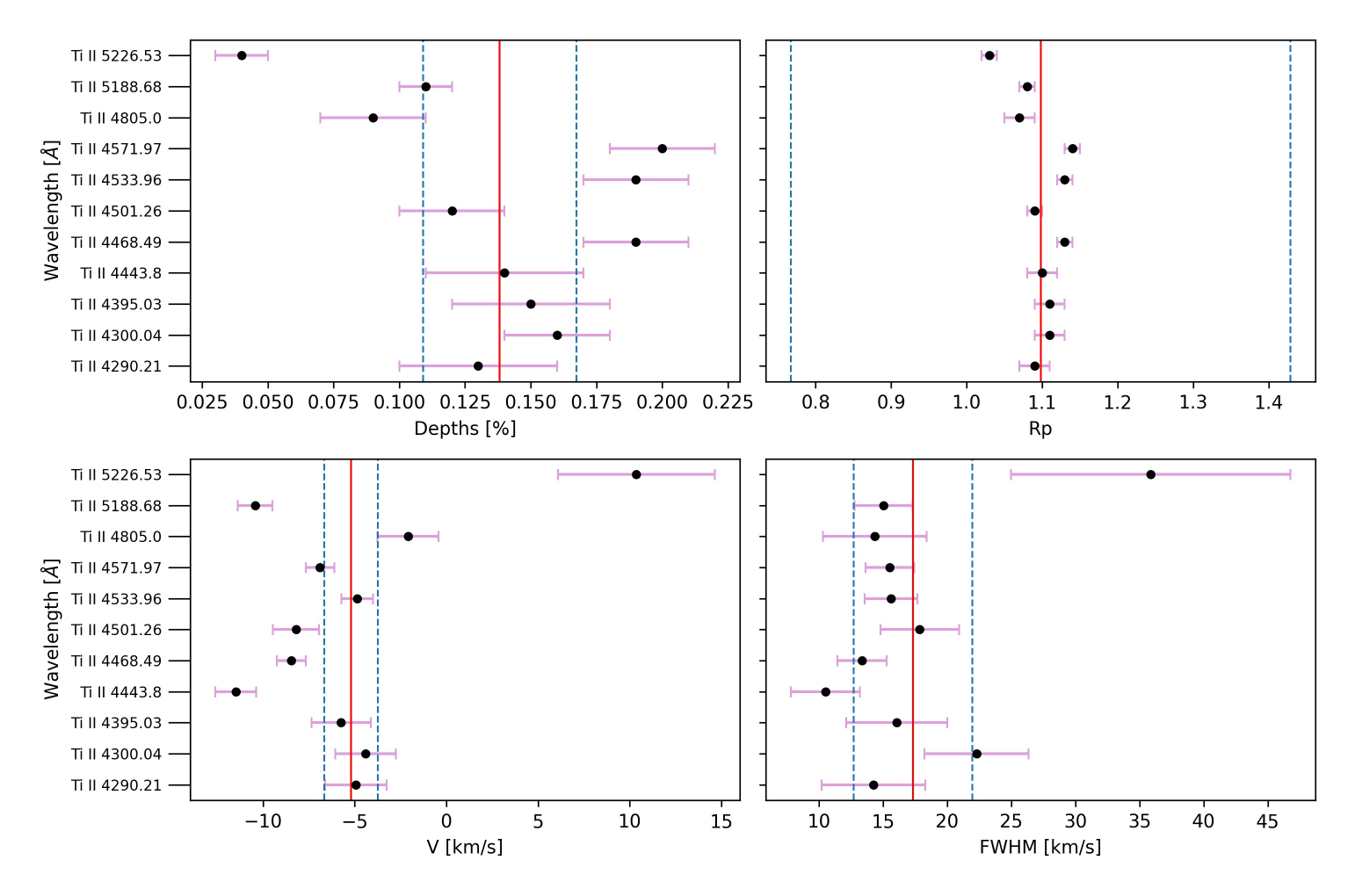}\quad
\caption{Best-fit parameters for each Ti{\sc ii} line. The red and dashed blue lines represent, respectively, the mean and its uncertainty.}
    \label{tiII_sp}
\end{figure*}

Starting from the NIST database, we searched for the strongest Ti lines in the wavelength range that we have available. We confidently detected ten Ti {\sc ii} lines with a significance greater than 3$\sigma$ (Table~A.2). We confirm the previous three Ti {\sc ii} detections made by \cite{cauley2019atmospheric} and uncover eleven new ones. Overall, we find an average line depth of 0.14$\pm$0.03$\%$ and an average FWHM of 17.34$\pm$4.61 km/s for the identified Ti {\sc ii} lines, which are comparable to previous results. Individual values for the detection significance, absorption depth, $R_{\mathrm p}$, FWHM, and $v_{\mathrm sys}$ are reported in Table~A.2 and shown in Fig.~A.10 and Fig.~\ref{tiII_sp}. To mitigate blending effects and avoid false detections, we excluded from the analysis some Ti {\sc ii} lines that were too close to other lines of different atomic species.

We could not identify Ti {\sc i}, despite it being tentatively detected by \cite{borsato2023mantis}, but with a lower significance ($\sim$3$\sigma$) even with cross-correlation, hinting at the weakness of the signal. Ti {\sc ii} has been previously detected via cross-correlation by \cite{hoeijmakers2018atomic, hoeijmakers2019spectral} and \cite{borsato2023mantis} with a significance of $\sim$18$\sigma$, $\sim$25$\sigma$, and $\sim$17$\sigma$, respectively, supporting our single-line detections. We notice a discrepancy between our resulting mean line shift from the zero point and those obtained by \cite{hoeijmakers2019spectral} and \cite{borsato2023mantis}. We recover a mean velocity shift of$-$5.19$\pm$1.45 km/s, while they reported systemic velocity values in the$-$18 to$-$20 km/s range. 
By subtracting the stellar systemic velocity that we adopted from their result (-17.74 km/s), we obtain $v_\text{sys}$ $\sim$ 0-3 km/s, and hence a difference in retrieved values, which may arise from the different methods and datasets used.

\subsubsection{Vanadium}
\label{sec:v}
Neither V {\sc i} nor V {\sc ii} were found by \cite{hoeijmakers2019spectral}, but recently V {\sc i} was tentatively detected by \cite{borsato2023mantis} at $\sim$3.5$\sigma$. We do not detect any V {\sc ii} single lines, but report the possible detection of one V  {\sc i} line at 4379.23 $\AA$ (4.3$\sigma$), as is shown in the tomography and in the transmission spectrum in Fig.~A.11. In a 2 $\AA$ range around the line of interest, we do not find possible contaminating species, supporting the detection of V {\sc i} in the atmosphere of KELT-9 b.

\subsubsection{Chromium}
\label{sec:cr}
Ionised chromium was successfully observed in previous studies in the atmosphere of KELT-9\,b via cross-correlation \citep{hoeijmakers2019spectral, borsato2023mantis} with a high statistical significance (>7$\sigma$). Neutral chromium was also detected at more than 5$\sigma$ by \cite{borsato2023mantis}, while \cite{hoeijmakers2019spectral} only report a tentative detection. Our single-line analysis reveals for the first time four Cr {\sc ii} lines with a significance larger than 3$\sigma$ (see Fig.~A.12 and Table A.2). Despite being detected by \cite{borsato2023mantis} via cross-correlation, we did not detect any neutral chromium line due to the weakness of the features.

\subsubsection{Iron}
\label{sec:fe}
\begin{figure*}
\sidecaption
\includegraphics[width=12cm]{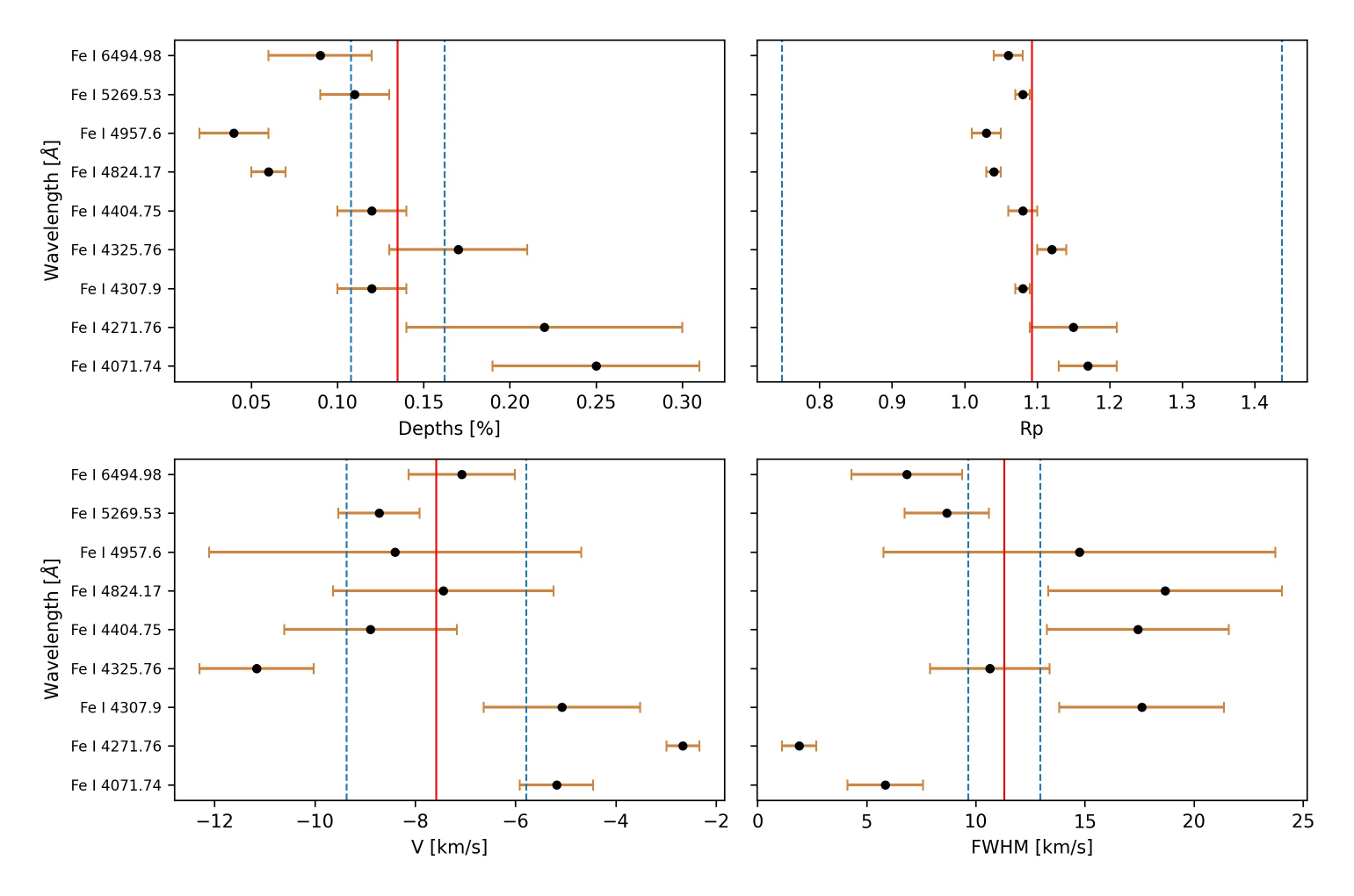}\quad
\caption{Same as Fig.~\ref{tiII_sp}, but for Fe {\sc i}.}
    \label{feI_med}
\end{figure*}

\begin{figure*}
\sidecaption
\includegraphics[width=12cm]{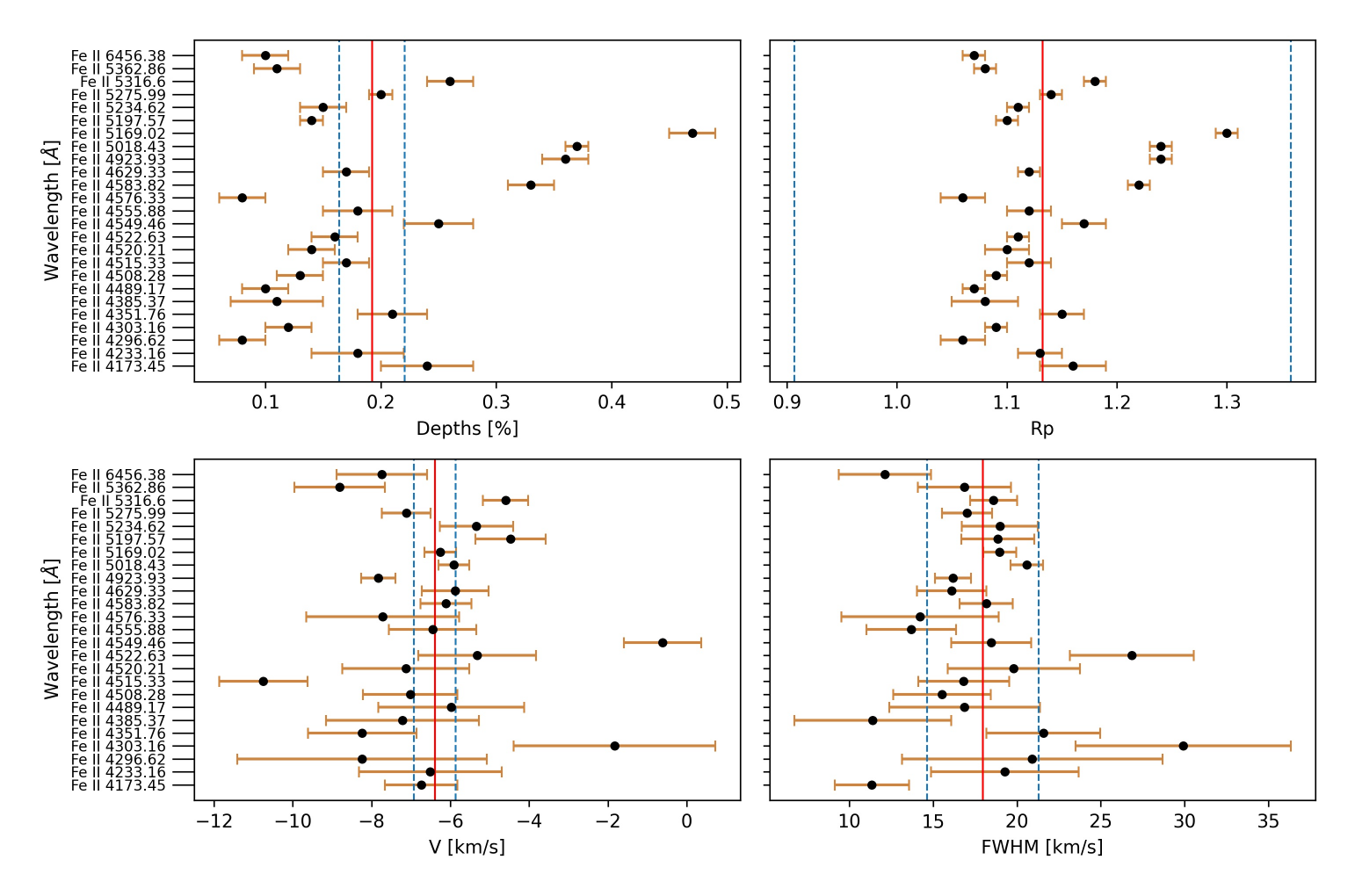}\quad
\caption{Same as Fig.~\ref{tiII_sp}, but for Fe {\sc ii}.}
    \label{feII_med}
\end{figure*}

Iron is the chemical species that displays the most lines in the optical range, and thus it has always been an ideal target for cross-correlation analysis. We focussed mainly on Fe {\sc i} and Fe {\sc ii}, which have been detected both via cross-correlation \citep{hoeijmakers2018atomic, hoeijmakers2019spectral, borsato2023mantis} and single-line analysis \citep{cauley2019atmospheric}, detecting 10 Fe {\sc i} and 25 Fe {\sc ii} lines. We also conducted an investigation into the presence of lines corresponding to Fe {\sc iii}, Fe {\sc iv}, and Fe {\sc v}; however, no detections were made in these cases. From Fig.~\ref{feI_med} and \ref{feII_med}, we can see that Fe {\sc i} lines are on average shallower and more narrow than Fe {\sc ii} lines. The velocity shift values of the Fe {\sc i} lines are in agreement among them, except for a few isolated points (which are still within 2$\sigma$). Instead, Fe {\sc ii} lines present a larger scatter in velocity shifts. The Fe {\sc i} and Fe {\sc ii} spectra are shown in Fig.~A.13 and Fig.~A.14, while the best-fit parameters are shown in Fig.~\ref{feI_med} and Fig.~\ref{feII_med}.

\section{Comparison with non-local thermodynamic equilibrium models}
\label{sec:comp_models}
The intense ultraviolet (UV) radiation that the planet experiences as a result of the host star's high temperature and close orbital separation leads to significant deviations from the local themodynamical equilibrium \citep[LTE;][]{fossati_models}. These authors highlighted how the H$\alpha$ and H$\beta$ transmission spectra are in an excellent match with their models once NLTE effects are taken into account in the computation of both the temperature-pressure structure and the transmission spectrum. Hence, the observations can be used to further constrain the theoretical models, and thus the physical properties of the exoplanet atmosphere. Furthermore, as \cite{fossati_models} mentioned, the NLTE synthetic transmission spectrum can be used to guide future observations aiming to detect features in the observed transmission spectrum, as has been done by \cite{borsa_ox_nature}. The NLTE transmission spectra considered in this work and published by \citet{fossati_models} were computed using the Cloudy for Exoplanets (CfE) interface, which enables one to use the Cloudy general-purpose NLTE radiative transfer code to model the atmospheric structure of middle and upper planetary atmospheres and generate transmission spectra. All necessary information about CfE, Cloudy, and the NLTE models of KELT-9b can be found in \citet{fossati_models}.

As for the results in Section \ref{sec:results}, we divided our comparison with NLTE models into two sections, one for the hydrogen Balmer series and one for the metal lines. \cite{fossati_models} and \cite{fossati_comp} already compared their NLTE models with those obtained by other works detecting Balmer lines mentioned in Section \ref{sec:H}.
For the analysis of the detected metal lines, we used the NLTE models to support our detections, checking the presence of features in the NLTE spectra in the same region in which we find an absorption signal in the observed transmission spectrum. To further verify that an absorption feature in the NLTE spectrum was due to a specific species, we also used NLTE models computed accounting only for hydrogen and the specific species considered. Hereafter, we refer to the latter as `individual NLTE models', while we call global NLTE models the NLTE transmission spectrum computed accounting for all species.

Then, we carried out an analysis similar to \cite{borsa_ox_nature},
whereby we used different microturbulence ($\nu_{\mathrm mic}$; velocity of gas on a scale smaller than the pressure scale height, implemented by adding it in quadrature to the thermal velocity) and macroturbulence ($\nu_{\mathrm mac}$; velocity of gas on a scale larger than the pressure scale height) velocity values to account for any additional line broadening. We created a grid of models with different $\nu_{\mathrm mic}$ values ranging between 1 and 14 km/s (in steps of 1 km/s), and for each model we convolved the transmission spectrum with: i) the rotational profile of the annulus representing the atmosphere (as done by \citealt{brogi_rot}), ii) a step function to include the broadening
due to exposure time (with a size of 4.7 km/s, retrieved from the in-transit mean exposure time of 404 s) and, iii) a Gaussian kernel with an FWHM equal to a $\nu_{\mathrm mac}$ broadening between 1 and 25 km/s, in steps of 1 km/s. We then selected the $\nu_{\mathrm mic}$--$\nu_{\mathrm mac}$ pair that minimises the $\chi^2$. To this end, for each spectral feature, we shifted the NLTE synthetic spectrum to match the centres of the observed lines. This was necessary because, when generating spectra, the Cloudy code \citep[used to generate the NLTE spectra;][]{2017RMxAA..53..385F} does not locate spectral lines at the expected wavelength, but at the centre of the spectral bin in which the line falls, which thus depends on the spectral resolution used to compute the synthetic spectra. This is why the NLTE synthetic spectrum cannot be used to derive line shifts from the observations. We remark that this coarse placement of spectral lines impacts exclusively the computation of the output spectra and not the NLTE radiative transfer, which is instead computed considering the actual line wavelengths. Before the comparisons, all the models are normalised by dividing for a fitted continuum in a small range around the line. The best-fitting NLTE synthetic spectra for each line are shown along with the observed transmission spectra and Gaussian fits from Fig.~\ref{balmer_sp} to A.14. The best $\nu_{\mathrm mic}$--$\nu_{\mathrm mac}$ values along with the $\chi^2_\text{red}$ values are listed in Table \ref{balmer_tab}. We stress that the $\chi^2_\text{red}$ values that we report in Table \ref{balmer_tab}, Table A.2, and Table A.3 may be underestimated due to the search for $\nu_{\mathrm mic}$--$\nu_{\mathrm mac}$ models that minimise $\chi^2_\text{red}$.

\subsection{Hydrogen Balmer series}
\cite{fossati_models} already presented a comparison between their NLTE and LTE synthetic spectra for H$\alpha$ and H$\beta$ and the transmission spectra presented by \cite{yan2018extended}, \cite{turner2020detection}, \cite{cauley2019atmospheric} and \cite{wyttenbach2020mass}, noticing a good $\chi^2$ agreement, and that the NLTE synthetic transmission spectrum fits the observations significantly better than the LTE model.
We extended their comparison for H$\alpha$ and H$\beta$ to our results including the other Balmer lines detected.
The $\chi^2_\text{red}$ reported in Table \ref{balmer_tab} suggests that NLTE models are in good agreement with our results: the H$\alpha$ and H$\beta$ $\chi^2_\text{red}$ values are compatible with the ones listed in \cite{fossati_models} and all the values are below 1 except for H$\zeta$. 
This discrepancy may be explained by the low S/N of the observations in that spectral region that gives rise to normalisation problems. However, despite the low $\chi^2_\text{red}$, the NLTE model predicts the existence of the H$\zeta$ line, supporting our detection.

\subsection{Metals}

For the analysis of the metal lines, it is important to consider that the NLTE models have been computed using solar abundances and that the synthetic transmission spectra have been computed using the sub-stellar temperature-pressure profile, which probably overestimates the temperature in the terminator region probed by the observations \citep{fossati_models}. These two effects may lead to discrepancies in the comparison with the observations in terms of the depth and shape of the lines.\\
- \textbf{Na}: The Na lines are both predicted by the NLTE models, which, however, seem to expect deeper lines, possibly due to a different abundance.\\
- \textbf{Mg}: The Mg  {\sc i} triplet b lines detected are predicted by the individual Mg {\sc i} NLTE model, which also confirms the presence of a Mg  {\sc i} line at 5528.40 $\AA$, corroborating our detections. We stress also that the best $\nu_{\mathrm mic}$--$\nu_{\mathrm mac}$ pair for this single line is in agreement with the ones found for the two reddest lines of the triplet.\\
- \textbf{Ca}: The synthetic NLTE spectra predict the existence of both the Ca {\sc i} lines that we detected, although these two lines are just above the $3\sigma$ threshold for detection. The Ca {\sc ii} H line shows a larger broadening than the Ca {\sc ii} K line, as was already found by the Gaussian fit, with a $\nu_{\mathrm mic}$ of 8 km/s and 4 km/s, respectively.\\
- \textbf{Sc}: The comparison with NLTE models and the use of individual species is particularly important for species like scandium that have never been detected via single-line analysis and that could be confused for other lines. In our case, the global NLTE model predicts the existence of all detected Sc {\sc ii} lines, but individual models show that the 5657 $\AA$ and 5684 $\AA$ lines might be blended with other weaker lines, respectively a close Sc {\sc ii} and Na I, with the latter line not detected in the atmosphere of KELT-9\,b. \\
- \textbf{Ti}: None of the detected lines are predicted by the individual Ti {\sc ii} NLTE models, while few of them may originate in the contamination of other species predicted by the global NLTE model. For the lines for which the global NLTE model does not predict the existence of any feature due to other species, we checked the presence of other lines of species detected in the VALD3 database \citep{vald}, which is more complete compared to NIST. We conclude that the detected features may be due to ionised titanium, which has been detected via single-line analysis and cross-correlation by \cite{hoeijmakers2018atomic, hoeijmakers2019spectral} and \cite{borsato2023mantis}.
\\
- \textbf{V}: Neither the global nor the individual NLTE models predict the presence of the V {\sc i} line at 4379.23 $\AA$ that we tentatively detected. The line is present in the NIST database and the reason why the NLTE model does not predict it may be related to the solar abundance adopted by the model or to the assumption on the temperature profile in the terminator region.\\ 
- \textbf{Cr}: As for scandium, the comparison with the NLTE model is particularly useful for the identification and detection of chromium lines; the individual Cr {\sc ii} synthetic spectrum predicts the existence of all detected lines showing also a blend with weaker lines (at 4558.64 $\AA$ and 4618.8 $\AA$) that seems not to affect our detections. Also, in this case, the individual NLTE model does not forecast the presence of any of the Cr {\sc i} features, supporting our non-detection. \\
- \textbf{Fe}: The synthetic NLTE models predict the presence of the detected Fe {\sc i} lines, except for the line at 4824.17 $\AA$ previously detected by \cite{cauley2019atmospheric}, which is instead due to Cr {\sc ii}. The large FWHM obtained for the line at 5328.03 $\AA$ is due to blending with two Fe {\sc i} lines, while the analysis of the 5269.53 $\AA$ line shows the presence of a strong nearby Ca {\sc i} line. Overall, we report a good match between our Fe {\sc ii} detections and the theoretical NLTE model. We remark that the Fe {\sc ii} lines at 5316.6 $\AA$ and 4351.76 $\AA$ might be affected by blending.

\section{Discussion}
\label{sec:discussion}
\begin{figure}
\includegraphics[width=\columnwidth]{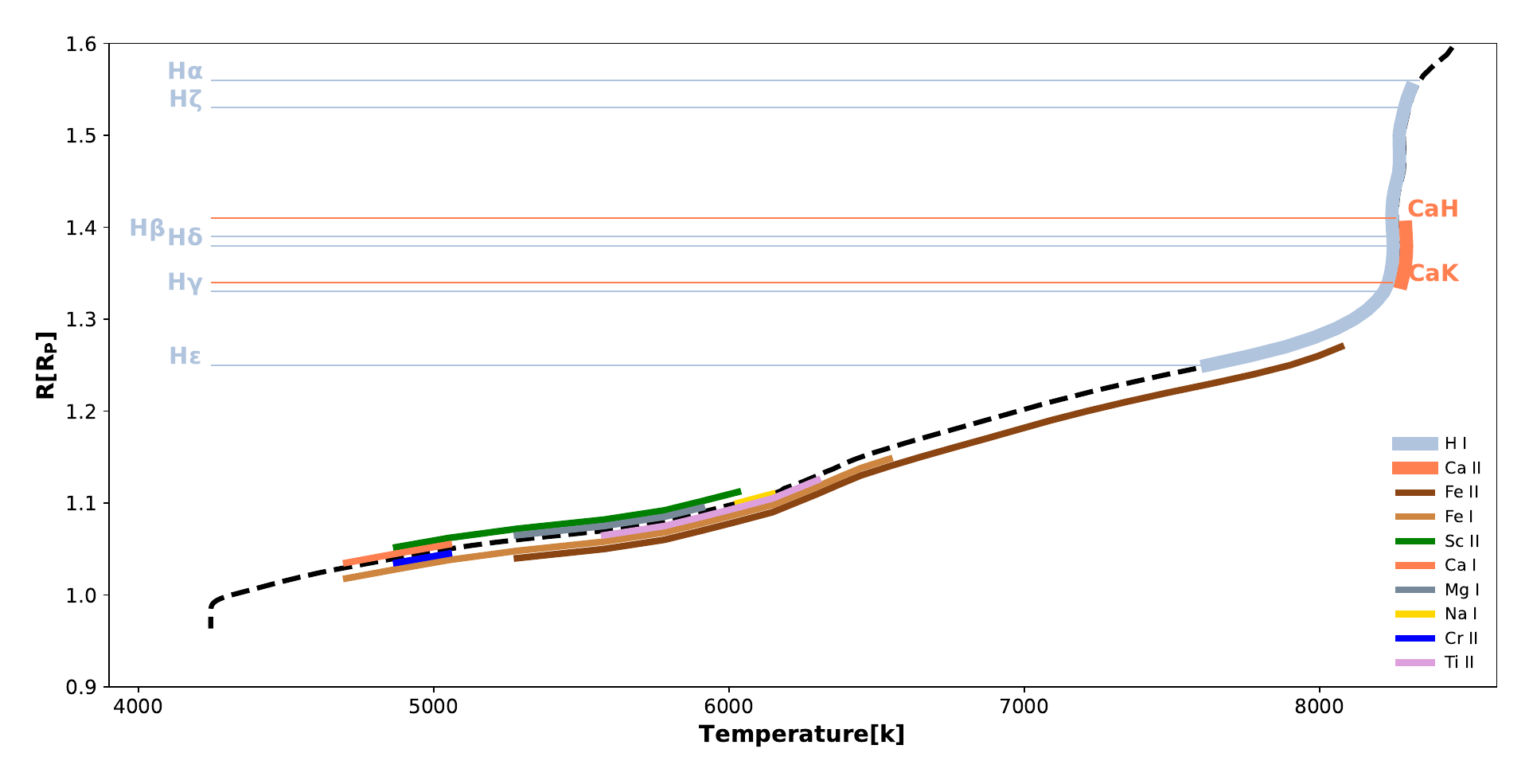}\quad
    \caption{Height distribution of chemical species detected expressed in planetary radii as a function of atmospheric temperature. H {\sc i} and Ca {\sc ii} lie on top, but still well below the Roche (1.95 $R_\text{p}$). The other metals are distributed at lower altitudes. The dashed black line represents the temperature profile presented by \citet{fossati_models} accounting for NLTE effects. The full lines representing each species height distribution are shifted with respect to the dashed black line for clarity.}
    \label{heights}
\end{figure}

\begin{figure}
\includegraphics[width=\columnwidth]{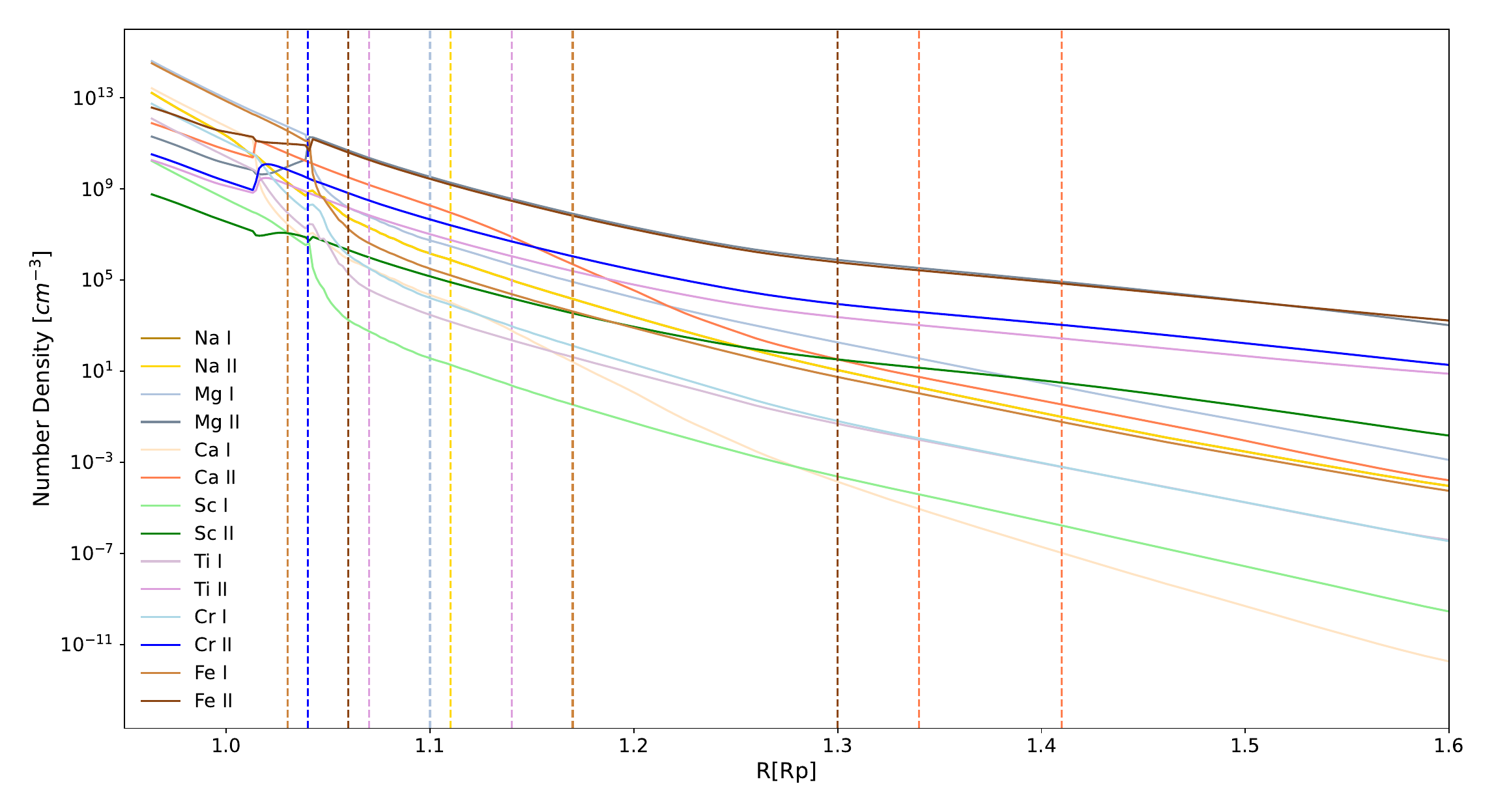}\quad
    \caption{Number densities predicted by the NLTE model of \citet{fossati_models} for the detected atoms in neutral and singly ionised form. The dashed lines indicate the height probed by the observations. Generally, the ionised species have higher number densities compared to the neutral ones.}
    \label{abundances}
\end{figure}

The large number of single lines detected allows us to better constrain the behaviour of KELT-9 b's atmosphere. In the analyses discussed below, we excluded the lines with large FWHMs that the comparison with NLTE models proved to be blended with other lines.
By coupling the depth of the line with the altitude at which it forms, we were able to construct a map of the species detected as a function of altitude. Our observations allowed us to probe the atmosphere up to 1.5 $R_\text{p}$, which is still well below the Roche lobe (1.95$R_\text{p}$). In the upper atmospheric layers, $\sim$1.25 $R_\text{p}$, neutral hydrogen and Ca {\sc ii} H\&K form, while ionised iron forms below, extending down to 1.1-1.0 $R_\text{p}$, where it mixes with neutral Fe. Just below 1.2 $R_\text{p}$ we find Ti {\sc ii}, while between 1.10 and 1.04 we find scandium and Cr {\sc ii}. Neutral calcium is the species that forms deeper in the atmosphere, just above 1.03 $R_\text{p}$.
Using the temperature-radius profile from \citet{fossati_models}, we can couple the heights that we retrieved from the Gaussian fits with the modelled atmospheric temperature obtained accounting for NLTE effects, as is shown in Fig.~\ref{heights}. We find that, in the upper atmosphere, H {\sc i} Balmer lines form at a temperature higher than 7500 K, while Ca {\sc ii} lies above 8200 K. 
Slightly below 1.3 Rp, we find Fe {\sc ii} ranging from 5000 K to 8000 K. Ti {\sc ii} and Fe {\sc i} show a common range in height corresponding to a temperature of 4800--6600 K. Sc {\sc ii} and Mg {\sc i} lie between 5000 K and 6000 K, while Ca {\sc i} and Cr {\sc ii}, which lie lower in the atmosphere, have temperatures below 5000 K.\\
The NLTE model's predicted number densities for each neutral and ionised species support our detections. As a matter of fact, the general trend is that ionised species have higher abundances compared to the neutral ones (see Fig.~\ref{abundances}), corroborating the detection of Sc {\sc ii}, Ti {\sc ii}, and Cr {\sc ii}. We are not able to find ionised magnesium due to the lack of strong lines in the HARPS-N wavelength range, while for Ca and Fe we detect both the neutral and the ionised lines, probably because of their higher abundance compared to the elements for which we find only ionised species.

\begin{figure*} 
\includegraphics[width=\columnwidth*2]{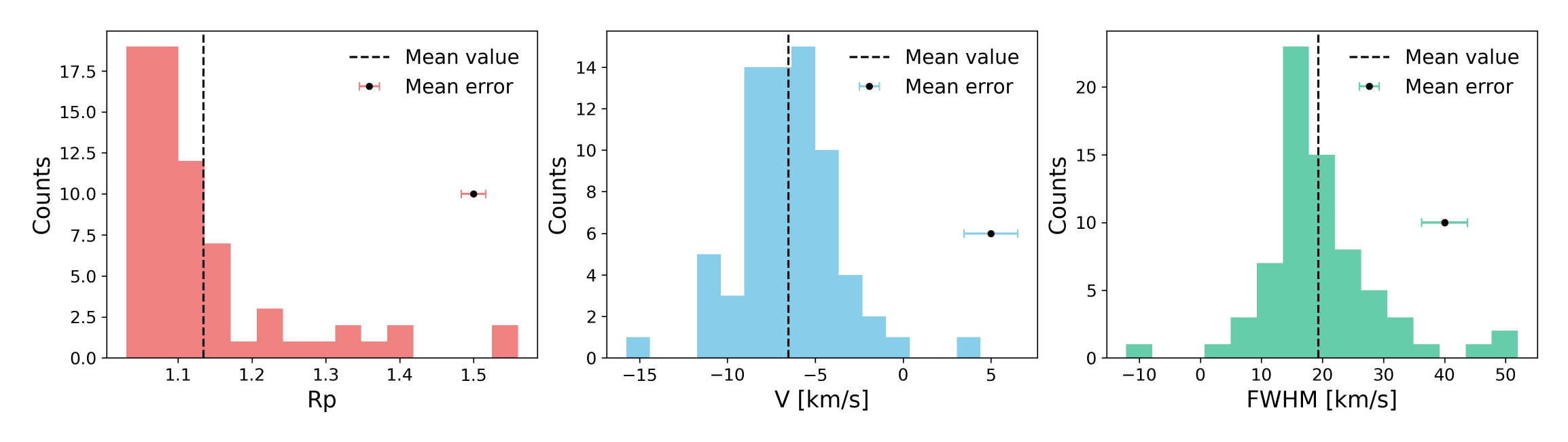}\quad \\
\includegraphics[width=\columnwidth*2]{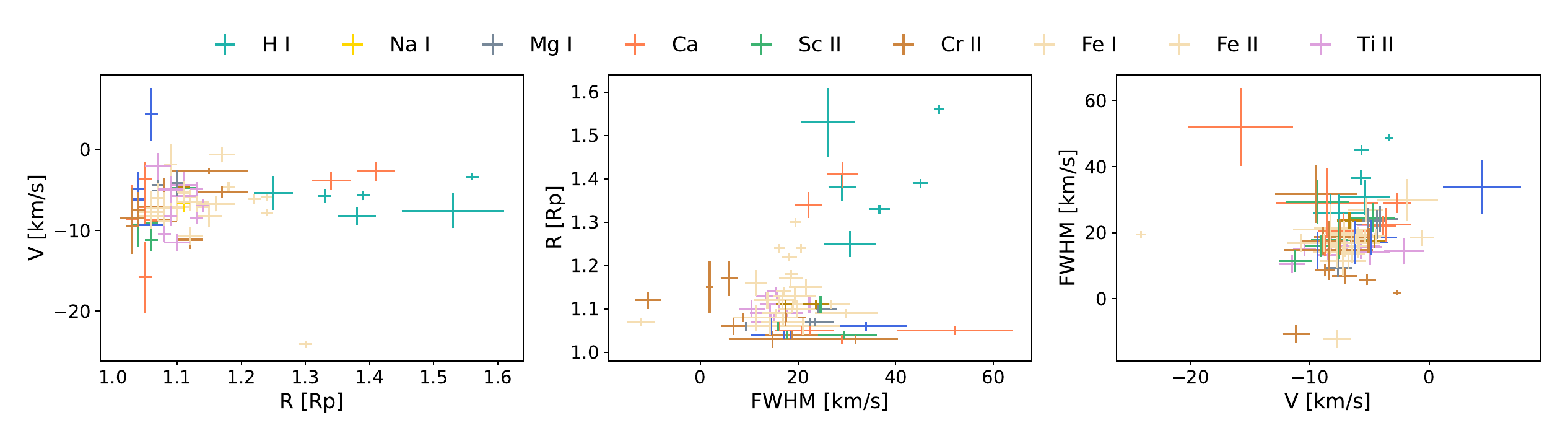}\quad \\
\includegraphics[width=\columnwidth*2]{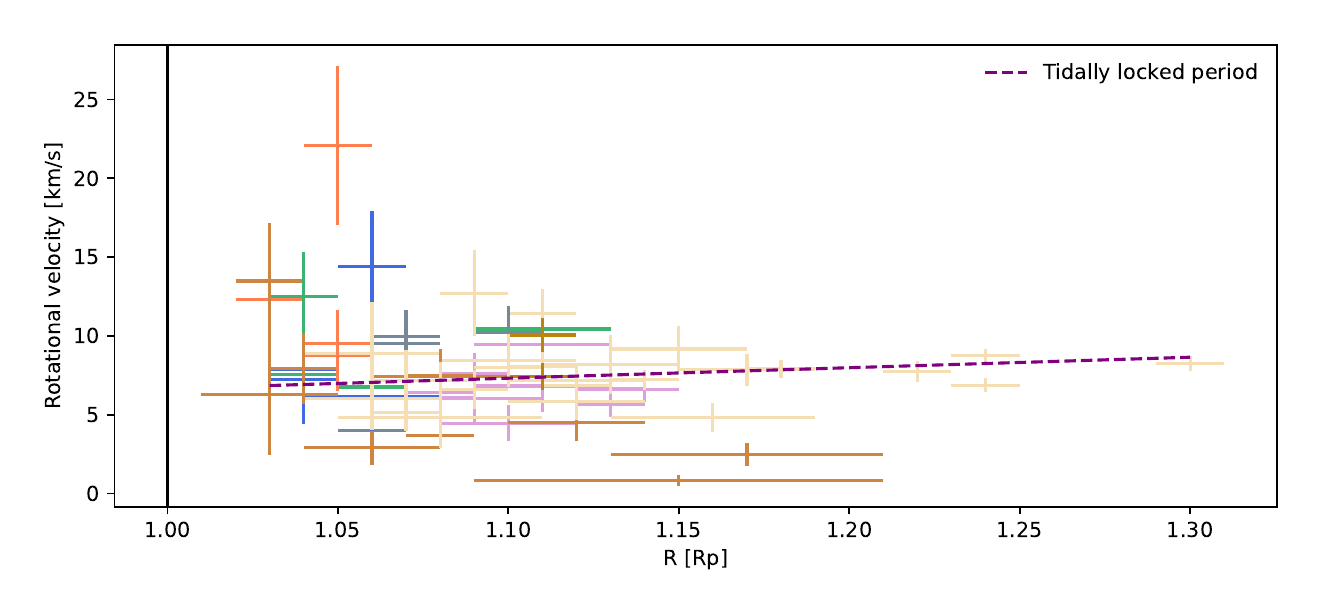}\quad
    \caption{Analysis of Gaussian best-fit parameters of all detected lines. \textit{Top panel}: Histograms of height, velocity, and FWHM. \textit{Middle panel}: Relations between velocity, height, and FWHM. \textit{Bottom panel}: Rotational velocity against the height in the atmosphere (in planetary radii); the dashed purple line represents the profile expected for the tidally locked scenario, while the black line represents the planetary radius.}
    \label{relations}
\end{figure*}

Fig.~\ref{relations} shows the possible relations among our Gaussian best-fit parameters. The distribution in height has already been discussed before, but from the histogram we can clearly see how most of the lines detected are below 1.15 planetary radii, while only a few lines of Fe {\sc ii}, Ca {\sc ii} and the Balmer lines lie above.
Overall, we find all lines to be blueshifted, with the distribution peaking around $-$6km/s and ranging from 0 km/s to $-$11 km/s, except for two points above 0 km/s. We do not find a clear correlation between the velocity shift and the height distribution, suggesting the homogeneous motion of the atmosphere in the probed layers. The systematic blueshift is expected by the presence of night-side to day-side winds due to the fact that we are observing the planet terminator. General circulation models for HJ predict wind speeds of typically 2--4 km/s, which is slightly smaller than the 6 km/s that we retrieve.
The FWHM of the lines detected returns a quasi-symmetric distribution peaked at $\sim$20 km/s, with a tail due to the Ca {\sc ii} and H Balmer lines, which have a different physical origin since their broadening is mostly due to collisional effects rather than rotational ones.

We focussed on the possible correlation between the FWHM and the height distribution, comparing the line widths and the height distribution in Fig.~\ref {relations} in a similar way to that done by \cite{borsa_rot}. We excluded the H {\sc i} and Ca {\sc ii} lines for the reason explained above. Assuming the planet and its atmosphere rotate to be a rigid body with a rotation period equal to 6.6 km/s (tidally locked), we obtained the velocity-radius profile shown by the purple line in Fig.~\ref{relations}. Despite most of our points being at lower radii, our results follow the tidally locked profile, supporting the expectation of tidally locked rotation. In our analysis, we retrieved the FWHM from a Gaussian fit. Although this does not take into consideration the fact that during the transit only an annulus is visible, and that another profile would be required \citep{brogi_rot}, we decided to fit a Gaussian profile as we aim for a qualitative comparison between the statistics inferred from our results and the expected rotational profile shown in purple in Fig.~\ref{relations}.

The impact of NLTE effects in the atmosphere of KELT-9\,b has already been discussed by various studies \citep{fossati2020data, fossati_models}, and the inclusion of NLTE effects in the modelling scheme has been found to be necessary to reproduce the observations. The key impact of NLTE effects in the planetary atmosphere is the significant increase in temperature in the middle and upper atmosphere. The overpopulation of Fe {\sc ii} drives heating, significantly increasing the absorption of stellar near-UV radiation, where the host star’s spectral energy distribution peaks, further increasing the heating rate. This process leads to the strong temperature inversion that we see in Fig.~\ref{heights}.

We compared our line fitting results to NLTE models, finding that the line profiles, including the line depths, are well described by the NLTE synthetic spectra. Furthermore, NLTE models have played a key role in identifying and confirming the single-line detections presented here, hinting at the blending of different lines, which allowed us to explain why few lines show large FWHM values.
We computed the mean $\nu_{\mathrm mic}$ and $\nu_{\mathrm mac}$ for each species, obtaining the results listed in Table \ref{vmic_tab}.
\begin{table}
\caption{Mean values of $\nu_{\mathrm mic}$ and  $\nu_{\mathrm mac}$ for each species detected.}
\centering
\begin{tabular}{lll}
\hline
 Species \tablefootmark{a}   &  $\nu_{\mathrm mic}$ [km/s] & $\nu_{\mathrm mac}$ [km/s]
\\
\hline
H {\sc i} & 3.67 $\pm$ 0.93 & 16.50 $\pm$ 4.12 \\
Na {\sc i} & 1.0 $\pm$ 0.02 & 13.0 $\pm$ 3.5 \\
Mg {\sc i} & 1.75 $\pm$ 0.12 & 19.50 $\pm$ 3.63 \\
Ca {\sc i} & 3.40 $\pm$ 0.61 & 15.6 $\pm$ 5.02 \\
Sc {\sc ii}& 4.50 $\pm$ 0.97 & 13.0 $\pm$  7.73 \\
Cr {\sc ii} & 2.00 $\pm$ 0.34 & 13.0 $\pm$ 4.91 \\
Fe {\sc i} & 2.70 $\pm$ 1.07 & 6.90 $\pm$ 4.54 \\
Fe {\sc ii} & 2.96 $\pm$ 0.28 & 7.92 $\pm$ 1.66 \\
\textbf{All the species} \tablefootmark{b} & 3.03 $\pm$ 0.25 & 10.03 $\pm$ 1.43 \\
O {\sc i} \tablefootmark{c} & 3.0 $\pm$ 0.7 & 13 $\pm$ 5 \\
\hline
\label{vmic_tab}
\end{tabular}
\tablefoot{\tablefoottext{a}{We excluded from the analysis the Ti II lines, since the NLTE models do not predict their presence. \tablefoottext{b} We did not include the H I and Ca II lines, since their broadening is mostly due to collisional effects.} \tablefoottext{c}{Results from \cite{borsa_ox_nature} with CARMENES data.}}
\end{table}

\begin{figure}
\includegraphics[width=\columnwidth]{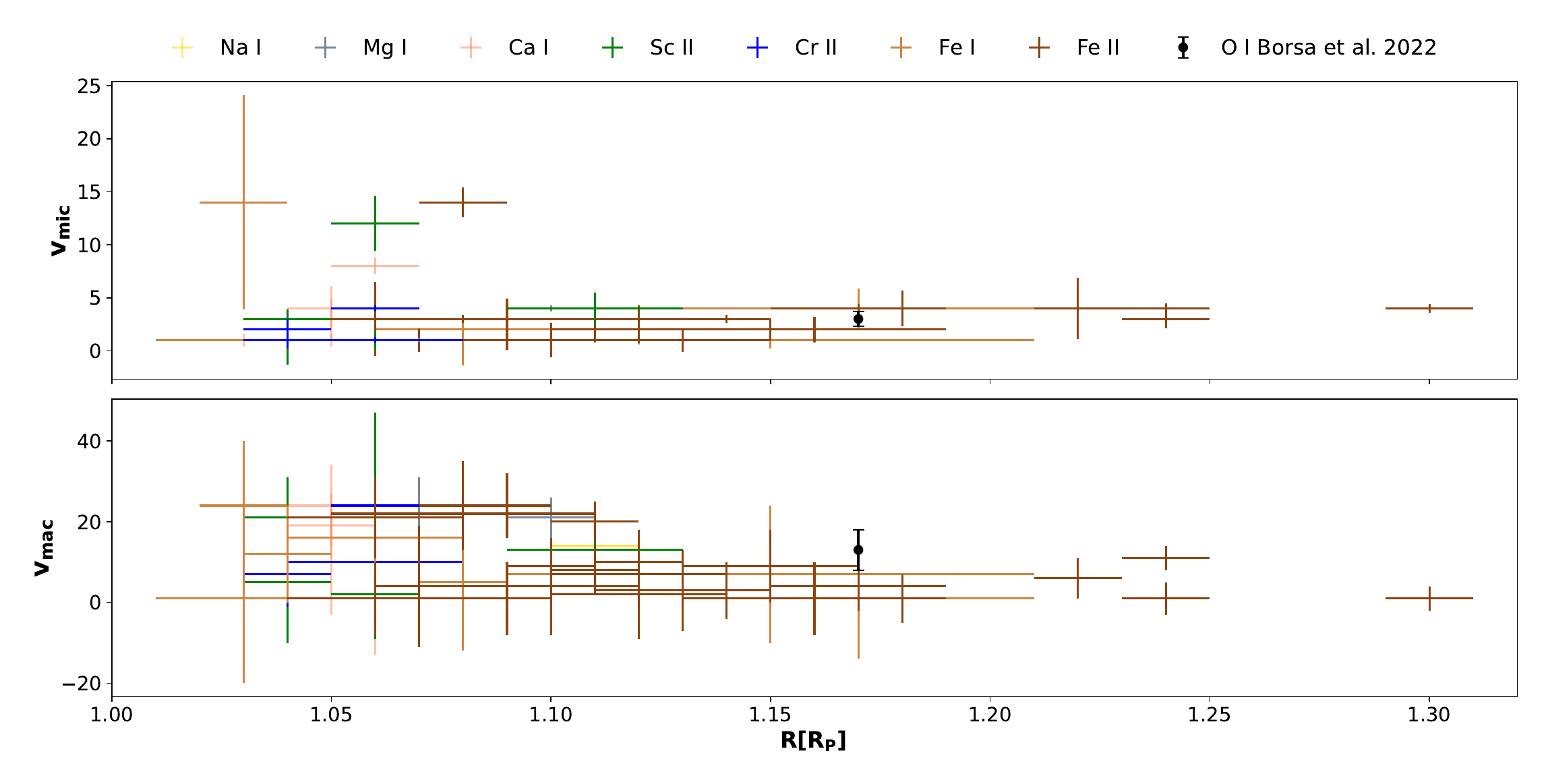}\quad
    \caption{Distribution of $\nu_{\mathrm mic}$ and $\nu_{\mathrm mac}$ as a function of height, for the different species detected. We also report, as a comparison, the values retrieved by \cite{borsa_ox_nature} with CARMENES data.}
    \label{vmic_fig}
\end{figure}

\cite{borsa_ox_nature} found $\nu_{\mathrm mic}$ = 3.0 $\pm$ 0.7 km/s and $\nu_{\mathrm mac}$ = 13 $\pm$ 5 km/s, analysing the oxygen triplet at about 7770 \AA\ with CARMENES data.
The mean values of $\nu_{\mathrm mic}$ and $\nu_{\mathrm mac}$ that we retrieve including all the species except for H {\sc i}, Ca {\sc ii}, and Ti {\sc ii} agree with their results.
We investigated the possible correlation between $\nu_{\mathrm mic}$ and $\nu_{\mathrm mac}$ and the height distribution, shown in Fig.~\ref{vmic_fig}, and we did not find any significant correlation.
We stress that we consider $\nu_{\mathrm mic}$ and $\nu_{\mathrm mac}$ to be fudge parameters needed to explain the observed extra broadening, but they still do not have a well-defined physical meaning. The $\nu_{\mathrm mic}$ values, considering their uncertainties, are compatible with zero.

\section{Conclusions}
\label{sec:conclusions}
In summary, we analysed six transit observations of the UHJ KELT-9\,b taken with the HARPS-N spectrograph mounted at the TNG. The elevated temperature of the planet results in the significant extension of the atmosphere and leads to an increase in the S/N of the data, making it the perfect target for transmission spectroscopy.
We employed the single-line analysis technique to search for a variety of metal lines across different species, ranging from the lightest elements, such as hydrogen, to the heaviest ones, such as iron. Thanks to the strong atmospheric signal due to the combination of six observational nights, we were able to identify 70 individual lines belonging to seven different species, Na {\sc i}, Ca {\sc i}, Ca {\sc ii}, Fe {\sc i}, Fe {\sc ii}, Mg {\sc i}, Ti {\sc ii}, Sc {\sc ii}, and Cr {\sc ii}. This is more than any other study utilising this technique has ever found. Our approach allowed us to confirm previous detections and make new ones, such as the discovery of H$\epsilon$ and H$\zeta$, which had only been deemed tentative up to now. H$\zeta$ displays a larger depth than expected and we stress that our results might be influenced by normalisation problems arising from the low S/N in the bluest part of the wavelength range covered by HARPS-N. We also detected Cr {\sc ii} and Sc {\sc ii} single lines for the first time in the atmosphere of an exoplanet, along with many new lines from already-detected species.
Among the chemical species that we looked for, we could not find He, Li, Be, K, Sr, Y, Ni, Ba, and Mn single lines. Even though the atmospheric signal is strong, with an average S/N of $\sim$100, one possible reason for these non-detections is the weakness of those lines, which might not be sufficient for a single-line analysis study. 

Single-line analysis in transmission has proven to be a powerful tool to support and confirm detections made with cross-correlation studies, while delivering important physical information about the width and depth of the lines, which in turn provides clues about the turbulence and the stratification of the atmosphere, in addition to information on the physical and chemical structure of the atmosphere. By juxtaposing our absorption lines with NLTE models, we could verify the presence of detected features and distinguish regions in which line blending may be occurring. Most of the lines analysed experience a velocity blueshift, corroborating the existence of night-side to day-side winds in the atmosphere, especially for the lines forming between 1.07 and 1.20 planetary radii. The rotational velocity and the height distribution generally agree with the hypothesis of a tidally locked rigid rotating body. In conclusion, our study marks a significant contribution to the application of the single-line analysis technique: we yielded an unprecedented number of individual lines, detecting many of them for the very first time in an exoplanetary atmosphere. The huge number of lines detected allowed us to further corroborate the presence of winds and to clarify the atmospheric stratification of KELT-9b.

\begin{acknowledgements}
The authors acknowledge financial contribution from PRIN INAF 2019 and from the European Union - Next Generation EU RRF M4C2 1.1 PRIN MUR 2022 project 2022CERJ49 (ESPLORA), PRIN MUR 2022 project PM4JLH (“Know your little neighbours: characterising low-mass stars and planets in  the Solar neighbourhood”) and PRIN MUR 2022 (project No. 2022J7ZFRA,  EXO-CASH).
The authors acknowledge the support of the ASI-INAF agreement 2021-5-HH.0 and the project “INAF-Astrofisica Fondamentale GAPS2".
AS acknowledges financial contribution from the University College London MAPS PGR Travel Grant 2021/22, the European Research Council under the European Unions Horizon 2020 research and innovation program (grant agreement No. 758892, ExoAI) and the European Unions Horizon 2020 COMPET program (grant agreement No. 776403, ExoplANETS A). Furthermore, AS is supported by a UKRI-STFC CDT-DIS studentship under project No. 541466. GMa acknowledges support from CHEOPS ASI-INAF agreement n. 2019-29-HH.0. Part of the research activities described in this paper were carried out with contribution of the Next Generation EU funds within the National Recovery and Resilience Plan (PNRR), Mission 4 - Education and Research, Component 2 - From Research to Business (M4C2), Investment Line 3.1 - Strengthening and creation of Research Infrastructures, Project IR0000034 – “STILES - Strengthening the Italian Leadership in ELT and SKA”. This work has made use of the VALD database, operated at Uppsala University, the Institute of Astronomy RAS in Moscow, and the University of Vienna.\\
We sincerely thank the Referee for their comments stressing some important factors and analyses that we did not consider initially and that we think improved the quality of the manuscript.\\
\textit{Author contributions.} MCD led the data analysis for this project with contributions from AS and FB. LF provided and contributed to the analysis of NLTE atmospheric models. MCD wrote the manuscript along with AS, FB, LF and GM. All authors discussed the results and commented on the draft.\\
\textit{Data availability.} An extended version of the manuscript with an Appendix including Tables A.1-A.3 and Figures A.1-A.14 can be found on Zenodo \url{https://zenodo.org/records/13348484}.
\end{acknowledgements}

%
%
\bibliographystyle{aa} 
\bibliography{kelt9} 

\end{document}